\documentclass[smallextended,12pt]{svjour3}

\usepackage[export]{adjustbox}
\usepackage[utf8]{inputenc}
\usepackage{mathpazo}
\usepackage{amsmath}
\usepackage[round]{natbib}
\usepackage[margin=1in]{geometry}
\usepackage{amssymb}
\usepackage{standalone}
\usepackage{graphicx}
\usepackage{overpic}
\usepackage{epstopdf}
\usepackage[hyperfootnotes=false,colorlinks=true,allcolors=black,urlcolor=blue,anchorcolor=blue]{hyperref}
\usepackage[capitalise]{cleveref}
\usepackage{isomath}
\usepackage[export]{adjustbox}
\usepackage{caption}
\usepackage[normalem]{ulem}
\usepackage[rgb,dvipsnames]{xcolor}
\usepackage{tikz}
\usepackage{bm}
\usepackage[caption=false]{subfig}
\usepackage{booktabs}
\usepackage[math]{cellspace}

\AtBeginDocument{
    \heavyrulewidth=.08em
    \lightrulewidth=.05em
    \cmidrulewidth=.03em
    \belowrulesep=.65ex
    \belowbottomsep=0pt
    \aboverulesep=.4ex
    \abovetopsep=0pt
    \cmidrulesep=\doublerulesep
    \cmidrulekern=.5em
    \defaultaddspace=.5em
}

\definecolor{MyBlue}{rgb}  {0.1,0.1,0.9}
\definecolor{MyRed}{rgb}   {0.9,0.1,0.1}
\definecolor{MyGreen}{rgb} {0.05,0.4,0.05}
\definecolor{burntorange}{rgb}{0.8, 0.33, 0.0}
\definecolor{NeilMagenta}{rgb}{0.8, 0.1, 0.8}

\newcommand{\rev}[1]{#1}

\DeclareSymbolFont{otone}{OT1}{pplx}{m}{n}
\DeclareMathSymbol{\Delta}{\mathalpha}{otone}{1}

\crefformat{footnote}{#2\footnotemark[#1]#3}

\newcommand{\pd}[2]{\frac{\partial #1}{\partial #2}}
\newcommand{\pdd}[2]{\frac{\partial^2 #1}{\partial #2^2}}
\newcommand{\pdn}[3]{\frac{\partial^{#3} #1}{\partial #2^{#3}}}
\newcommand{\vnabla}{\boldsymbol{\nabla}}

\renewcommand{\vec}[1]{\boldsymbol{#1}}
\newcommand{\dx}{\Delta{x}}
\newcommand{\dy}{\Delta{y}}
\newcommand{\dt}{\Delta{t}}
\newcommand{\domain}{\Omega}

\newcommand \beq{\begin{eqnarray}}
\newcommand \eeq{\end{eqnarray}}
\newcommand \beqno{\begin{eqnarray*}}
\newcommand \eeqno{\end{eqnarray*}}
\newcommand \bit{\begin{itemize}}
\newcommand \eit{\end{itemize}}

\newcommand{\hereVPDE}[2][here]{\href{#2}{#1}\footnote{\url{#2}}}
\newcommand{\hereVPDEfig}[2][here]{\href{#2}{#1}\footnotemark{}}
\newcommand{\hereVPDELabel}[3][here]{\href{#2}{#1}\footnote{\url{#2}\label{#3}}}

\begin{document}
\title{VisualPDE: rapid interactive simulations of partial differential equations}
\titlerunning{VisualPDE: rapid interactive simulations of PDEs}
\author{Benjamin J.\ Walker \and Adam K.\ Townsend \and Alexander K.\ Chudasama \and Andrew L.\ Krause\footnote{Corresponding author \email{andrew.krause@durham.ac.uk}}}
\authorrunning{Walker et al.}

\institute{
    A. L. Krause, A. K. Townsend, A. K. Chudasama
    \at Department of Mathematical Sciences, Durham University, Upper Mountjoy Campus, Stockton Road, Durham DH1 3LE, United Kingdom \\
    B. J. Walker
    \at Department of Mathematical Sciences, University of Bath, Bath BA2 7AY, United Kingdom; Department of Mathematics, University College London, London WC1E 6BT, United Kingdom
    }

\date{Received: date / Accepted: date}

\maketitle

\begin{abstract}
Computing has revolutionised the study of complex nonlinear systems, both by allowing us to solve previously intractable models and through the ability to visualise solutions in different ways. Using ubiquitous computing infrastructure, \rev{we provide a means to} go one step further in using computers to understand complex models through instantaneous and interactive exploration. This \rev{ubiquitous infrastructure} has enormous potential in education, outreach and research. Here, we present VisualPDE, an online, interactive solver for a broad class of 1D and 2D \rev{partial differential equation (PDE)} systems. Abstract dynamical systems concepts such as symmetry-breaking instabilities, subcritical bifurcations and the role of initial data in multistable nonlinear models become much more intuitive when you can play with these models yourself, and immediately answer questions about how the system responds to changes in parameters, initial conditions, boundary conditions or even spatiotemporal forcing. Importantly, VisualPDE is freely available, open source and highly customisable. We give several examples in teaching, research and knowledge exchange, providing high-level discussions of how it may be employed in different settings. This includes designing web-based course materials structured around interactive simulations, or easily crafting specific simulations that can be shared with students or collaborators via a simple URL. We envisage VisualPDE becoming an invaluable resource for teaching and research in mathematical biology and beyond. We also hope that it inspires other efforts to make mathematics more interactive and accessible.
\end{abstract}
 

\maketitle

\section{Introduction}
This paper introduces VisualPDE, a web-based tool for interactive simulations of time-dependent partial differential equations \rev{(PDEs)} in one and two spatial dimensions. We would highly encourage a reader to browse to \url{https://visualpde.com/} to directly experience this tool, including exhaustive documentation and direct access to all code used. For each of the examples and figures presented in this paper, we have included links \rev{to the VisualPDE website in the form of clickable footnotes} so that the reader can interact with the content, live in their browser. Rather than provide detailed documentation of VisualPDE's features and examples (which can both be better experienced on the website), this paper provides wider context to the design and anticipated use cases of this tool. \rev{Our discussion here} includes aspects of the overall design philosophy, technical achievements, and how we imagine it being used for teaching, research, knowledge exchange, and outreach activities involving partial differential equations.

We first give broad historical, pedagogical, and technical context to this project in \cref{sec:history}. In \cref{sec:maths} we give a high-level overview of the technical frameworks used, with a particular eye towards demonstrating why key design choices were made, and in what ways others can extend this project or develop similar tools using the technologies underlying VisualPDE. Our user interface design is a key aspect of this work, described in \cref{sec:interaction} along with ways to share and reuse VisualPDE. In \cref{sec:examples} we give examples of using this tool within our own teaching, research, and outreach activities, suggesting ways it can be incorporated by others in future activities. We encourage reader\rev{s} to peruse these sections independently, and especially to spend time playing with the website, to get an overall feel for the project. Lastly, in \cref{sec:summary}, we summarise VisualPDE and our vision for how it may enhance how we communicate and interact with \rev{PDEs from diverse areas of mathematics}.

The website is intended to be living and constantly updated with new features and examples. As a result, figures and snapshots in this manuscript may not always reflect the precise content of the site, but should nevertheless remain illustrative of VisualPDE, its scope, and the authors' aims and ideas. Similarly, pages linked to via the URLs provided in this manuscript may evolve slightly over time, but we do not envisage this impeding the reader. 

\section{Historical \& pedagogical context}\label{sec:history}
Widespread access to computation has had many profound impacts on science, and society more broadly. The development of numerical scientific modelling, and the implementation of models on increasingly sophisticated hardware, has led to enormous advances across science and engineering \citep{gustafsson2018scientific}. Even relatively advanced and technical mathematical models, such as partial differential equations, have become widely used in numerous fields in part due to the development of  numerical methods \citep{thomee2001finite} and to the development of high-performance and personal computing machines. Computing has vastly increased the sophistication of modelling, enlarging the kinds of systems that we can understand through simulation that are not amenable to pen-and-paper calculations.

In the past few decades, the field of `ubiquitous computing' has emerged  to describe the accessibility of computation in many aspects of life \citep{meshram2016survey}, with widespread use of smartphones and tablet computers being the most obvious examples. In contrast to high-performance scientific computing, the emphasis here is not on technical capabilities, but on accessibility and everyday usability. This is readily apparent in the changing landscape of mobile and web computing, where we see an increasing emphasis on user experience \citep{benyon2019designing}, in addition to the availability of raw compute power in mobile devices. 

Some areas of scientific computing have transitioned to embracing this `ubiquitous' nature, with increasingly widespread use of online tools by students and educators for computation and visualisation. Importantly, it has also led to the development of more dynamic and interactive learning environments. For example, there is a growing literature on the use of web-based mathematical visualisation software applied to a variety of mathematical concepts. Such interactive and accessible tools aide students in being able to explore ideas on their own, including designing their own exercises and solutions, which has been shown to have substantial pedagogical benefits \citep{korucu2018effect}. Examples of such tools include  GeoGebra \citep{sangwin2007brief, arbain2015effects}, Desmos \citep{ebert2014graphing,king2017using}, and WolframAlpha \citep{dimiceli2010teaching, necesal2012experience}. Related to this are more computationally-focused web tools such as CodeRunner \citep{lobb2016coderunner} for rapid unit testing of student-written programs, as well as the growing use of Jupyter Notebooks \citep{cardoso2019using} for exploring a variety of different areas of scientific computing in interactive ways. Jupyter Notebooks have even been developed to explore areas of computational fluid dynamics \citep{castilla2023jupyter} and reaction--transport processes \citep{golman2019set}, showcasing the versatility of these web-based tools even for complex scientific computing tasks such as solving partial differential equations. We direct the interested reader to the work of \citet{engelbrecht2020transformation} for a broad overview on the role of internet technologies in transforming educational environments.

A major technical roadmark that enabled more immersive web-based tools was the transition from early web multimedia (e.g.\ Flash and Java applets) to more browser-native HTML5 `canvas' elements \citep{fulton2013html5} and related technologies. \rev{An example of such a technology-enabled website is `Complexity Explorables' \citep{complexity-explorables}, a collection of interactive simulations covering a wide range of topics in cellular automata, complex network theory, and beyond.} \rev{Another advance in this area is the release of} a range of graphics-oriented libraries that enable high level, platform agnostic development of interactive web-based applications that make use of graphics processing units (GPUs) to do large-scale calculations in real time \citep{angel2014interactive}, typically targeting 2D and 3D graphics. Libraries based on the popular WebGL framework include Three.js \citep{dirksen2013learning}, Babylon.js \citep{catuhe2014babylon} and Abubu.js \citep{kaboudian2019large}, which each allow for high-level interaction with GPUs within the context of a webpage and have been used in a variety of physics education scenarios \citep{mccauley2017browser,zatarain2023experiences}. Abubu.js in particular has been used to develop a range of mathematical visualisations, with a particular focus on models of cardiac electrophysiology \citep{kaboudian2021real, kaboudian2019real}. One can find a huge range of examples of this technology online by simply searching for `WebGL physics' and `WebGL fluid simulation'. 

Such tools have enabled crucial changes in the landscape of higher education instruction, particularly in the context of the life sciences. We note in particular the recent effort in teaching dynamical systems modelling to biology students by \cite{garfinkel2022teaching} through the development of a new course `Calculus for Life Sciences', based on dynamical systems. This course makes extensive use of Python/Jupyter Notebooks, allowing for an accessible and interactive approach to student-driven instruction. As described by \cite{garfinkel2022teaching}, there has been a growing need to modernise many aspects of undergraduate teaching, particularly in light of the rapid pace of technological and scientific advancement. This is especially true in the life sciences, where the disconnect with the more `classical mechanics' training of the mathematical, physical and engineering sciences has grown with the development of these fields \citep{woodin2010vision}. \rev{Teaching across and encouraging interaction between} the biological sciences \rev{and mathematical modelling}  also poses unique challenges in terms of subject matter and cultural differences \rev{within each field} \citep{reed2004mathematical}. Jupyter Notebooks and other high-level programming interfaces are an important aid to overcoming these barriers, though we feel that even more can be done to connect modelling work in the mathematical and physical sciences more directly with everyday experiences in the life sciences.

Interactive and accessible simulations can play an important role in bridging this divide by providing immediate access to application-relevant simulations, without the need to become experts in underlying foundational aspects (e.g.~modelling, numerical analysis and especially programming). In particular, we believe that there would be significant merit to a platform that abstracts away the intricacies of numerical methods and their implementation, with a user thereby free to play directly with concepts and models that are typically only accessible through simulation. Of course, depending on the goal of the learner, such a platform could also be used as a way to motivate learning about the more foundational topics rather than completely circumventing them. From this perspective, accessible and interactive simulations can provide a crucial tool for people to gain an understanding of more advanced topics without the need to build an expert knowledge of this from scratch. These ideas have already been pioneered in teaching coding skills by using `unplugged programming' methods that do not employ computers or traditional code \citep{sun2021comparing, munasinghe2023unplugged}.

VisualPDE aims to be such a platform, representing a significant advancement on the state-of-the-art of visualising PDE systems, especially those of relevance to modelling in mathematical biology and related fields. Existing solutions described above (e.g.\ Jupyter Notebooks or Mathematica) either still look and feel like programming, or they have a more point-and-click interface but can only handle a limited class of problems. In contrast, VisualPDE uses the extensible interactive design language of websites like Desmos, and applies this to PDE visualisation. It does this with no requirement for familiarity with numerical analysis or programming and, in addition, allows for an unprecedented range of freedom in terms of PDE systems, boundary and initial conditions, and other complex modelling and visualisation features. In addition, it is open source and by default coupled with a tutorial library that serves as a guide for any user looking for further instruction. In short, VisualPDE's accessible interface bridges the gap between the online software many students are (increasingly) familiar with, and more powerful methods of visualising solutions that typically have a higher barrier for entry.

In the next two sections, we will describe aspects of the implementation and design of VisualPDE, informed by the context described above. We will return to questions of using the software in education, research, and knowledge exchange in \cref{sec:examples}, demonstrating the impact that VisualPDE can have in these arenas.

\section{Implementing VisualPDE}\label{sec:maths}
VisualPDE is designed to be a flexible, plug-and-play PDE solver that runs in a web browser on a user's device. In this section, we give an overarching description of the equations that VisualPDE can solve, the numerical methods that underlie this and the aspects of the implementation that enable this to happen rapidly and interactively on widely available computing devices, including mobile phones.

\subsection{The PDEs}\label{sec:which-pdes}
When first designing VisualPDE, we were motivated by reaction--diffusion systems of the form
\begin{subequations}
\begin{align}
    \pd{u}{t} &= \vnabla \cdot(D_u\vnabla u) + f_u,\\
    \pd{v}{t} &= \vnabla \cdot(D_v\vnabla v) + f_v,
\end{align}
\end{subequations}
where $u,v$ are scalar fields defined on a domain $\domain\subset\mathbb{R}^2$, $D_u,D_v$ are given diffusion coefficients (often constants in classical applications) and $f_u,f_v$ are given functions of $u$ and $v$. An example of such a system is the \hereVPDE[Gray--Scott model]{https://visualpde.com/nonlinear-physics/gray-scott.html} \citep{gray1984autocatalytic}. This model received huge interest from scientists and artistic amateurs alike following numerical experiments by \cite{pearson1993complex}, which demonstrated a striking range of spatial and spatiotemporal phenomena by changing only two parameters. We note in particular the interactive WebGL simulators \citep{pmneila2012grayscott, karlsims}, among others, that served as  inspiration for VisualPDE. Our initial goal was to allow the user to type in values of the diffusion coefficients and kinetics and, hence, explore a larger class of reaction--diffusion systems, rather than hand code the WebGL as in these cited examples.

From this simple yet dynamically rich beginning, VisualPDE has been significantly extended. Currently, \rev{coupled systems or subsystems of four unknowns $(u,v,w,q)$} of the following general form can be posed in VisualPDE:
\begin{equation}\label{eq:general form}
\begin{aligned}
\pd{u}{t} &= \vnabla \cdot(D_{uu}\vnabla u+D_{uv}\vnabla v+D_{uw}\vnabla w+D_{uq}\vnabla q) + f_u,\\
\text{one of}\left\{\;\begin{aligned}\displaystyle\pd{v}{t} \\ v\end{aligned}\right. & 
\begin{aligned}
    &= \vnabla \cdot(D_{vu}\vnabla u+D_{vv}\vnabla v+D_{vw}\vnabla w+D_{vq}\vnabla q) + f_v \vphantom{\displaystyle\pd{v}{t}}, \\
    &= \vnabla \cdot(D_{vu}\vnabla u+D_{vw}\vnabla w+D_{vq}\vnabla q) + f_v,
\end{aligned}\\
\text{one of}\left\{\;\begin{aligned}\displaystyle\pd{w}{t} \\ w \end{aligned}\right. & 
\begin{aligned}
    &= \vnabla \cdot(D_{wu}\vnabla u+D_{wv}\vnabla v+D_{ww}\vnabla w+D_{wq}\vnabla q) + f_w  \vphantom{\displaystyle\pd{w}{t}}, \\
    &= \vnabla \cdot(D_{wu}\vnabla u+D_{wv}\vnabla v+D_{wq}\vnabla q) + f_w,
\end{aligned}\\
\text{one of}\left\{\;\begin{aligned}\displaystyle\pd{q}{t} \\ q\end{aligned}\right. & 
\begin{aligned}
    &= \vnabla \cdot(D_{qu}\vnabla u+D_{qv}\vnabla v+D_{qw}\vnabla w+D_{qq}\vnabla q) + f_q \vphantom{\displaystyle\pd{q}{t}}, \\
    &= \vnabla \cdot(D_{qu}\vnabla u+D_{qv}\vnabla v+D_{qw}\vnabla w) + f_q.
\end{aligned}
\end{aligned}
\end{equation}
\rev{Each of the variables $v$, $w$ and $q$ can either satisfy their own time-dependent PDE or can be specified directly in terms of the other variables.} The elements ${D}_{ij}$ form a matrix of diffusivities,  and the $f_i$ are given functions. Moreover, $f_i$ and ${D}_{ij}$ can be a function of the unknowns, the coordinates of the spatial domain, time, and any user-defined parameters. The functions ${f}_i$ can also depend on first and second spatial derivatives of $\vec{u}$ \rev{(mixed spatial derivatives are not currently supported)}. This flexibility in form entails that VisualPDE is able to solve a broad class of differential--algebraic equations, including systems with nonlinear cross diffusion and advection. By exploiting zeros and algebraic variables, one can construct systems with up to 8th order spatial or 4th order temporal derivatives. Many intricate, highly nonlinear PDEs can be cast in this general form. \rev{Examples on VisualPDE include} the \hereVPDE[inhomogeneous wave equation]{https://visualpde.com/basic-pdes/inhomogeneous-wave-equation.html}
\rev{\begin{equation}\label{eq:inhomog wave}
\pdd{u}{t} = \vnabla \cdot (f(x,y)\vnabla u)\quad \iff \quad \left\{\begin{aligned}&\pd{u}{t} = v,\\ &\pd{v}{t}=  \vnabla \cdot (f(x,y)\vnabla u), \end{aligned}\right.
\end{equation}}
and \hereVPDE[Keller--Segel chemotaxis models]{https://visualpde.com/mathematical-biology/keller-segel.html} \citep{horstmann20031970}
\rev{\begin{equation}\begin{aligned}\pd{u}{t}&=\nabla^2 u-\vnabla \cdot(\chi(u)\vnabla v)+f_u(u),\\ \pd{v}{t}&=D\nabla^2v+ f_v(u,v).\end{aligned}
\end{equation}}

We remark that the \rev{user interface (UI)} allows a user to select how many equations to solve, whether or not any of the cross diffusion terms ($D_{ij}$ for $i \neq j$) appear, and how many algebraic variables there are. In these cases the notation simplifies to represent the simpler versions of the system, including a single subscript in the case without cross diffusion. Similarly, a user may relabel any of the unknown variables to suit their preferred notation.

\subsection{Numerical methods}\label{sec:numerical-methods}
To solve equations cast in the general form of \cref{eq:general form}, VisualPDE employs a central finite difference scheme for all spatial derivatives \rev{(as well as some non-centred discretisations  to accommodate `upwinding' methods)} coupled to one of several explicit timestepping schemes. This non-specialised approach reflects the intended generality of VisualPDE and facilitates its plug-and-play features. A natural caveat of this generality is that some features of special systems (such as those with conserved quantities) may not be captured as well as might be achieved with bespoke numerical methods, such as symplectic or geometric integrators for models with symmetries such as Hamiltonian systems \citep{hairer2006geometric}. Nevertheless, we have found that the implemented scheme successfully captures key solution behaviours for a diverse range of systems currently implemented on the site, including many sensible approximations to systems with infinitely many conserved quantities. A nontrivial example is the phenomenon of solitons passing through one another in the \hereVPDE[Korteweg--De Vries equation]{https://visualpde.com/nonlinear-physics/kdv.html} \citep{miura1976korteweg}, 
\rev{\begin{equation}
\pd{\phi}{t}=-\pdn{\phi}{x}{3}-6\phi \pd{\phi}{x} \quad \iff \quad \left\{\begin{aligned}
     \pd{\phi}{t} &= - \pdd{v}{x} -6 v \phi,\\
    v& =  \pd{\phi}{x}.
    \end{aligned}\right.
\end{equation}
The use of explicit, non-symplectic timestepping means that this scheme will not preserve any of the infinitely many conserved quantities of this model and, hence, will exhibit small fluctuations as it is only an approximation to a `true' soliton-soliton solution. Nevertheless, after passing through one another, these solitons have visually identical heights and speeds in VisualPDE, indicating a good approximation of the behaviour of this model.}

The remainder of this subsection is a discussion of well-known standard finite difference methods, which we include for completeness. See \citet{leveque2007finite} or most other texts on the numerical analysis of partial differential equations for a more complete overview.

\subsubsection{Spatial discretisation}
In more detail, suppose that a user has taken the domain to be rectangular (the default for the majority of examples present on the site), with coordinates $(x,y)$ in $\domain=[0,L_x]\times[0,L_y]$ for $L_x,L_y>0$. For a given spatial step size $\dx$, configurable by the user, we split the domain into an $(L_x/\dx, L_y/\dx)$ grid (rounding down where necessary). Spatial derivatives are computed on this grid using commonplace finite difference schemes, with first derivatives being computed using the central difference:
\begin{equation}
    \left.\pd{u}{x}\right\rvert_{(t,x,y)} \approx \frac{u(t,x+\dx,y) - u(t,x-\dx,y)}{2\,\dx}.
\end{equation}
 The divergence terms of \cref{eq:general form} are computed by approximating
\begin{align}
     \left.\vnabla \cdot(D_u\vnabla u)\right\rvert_{(t,x,y)} \approx \frac{1}{2\,\dx^2} &\{D_u(t,x,y)[u(t,x-\dx,y) - 2u(t,x,y) + u(t,x+\dx,y)] \notag\\
    &+D_u(t,x-\dx,y)[u(t,x-\dx,y) - u(t,x,y)]\notag\\
    &+D_u(t,x+\dx,y)[u(t,x+\dx,y) - u(t,x,y)]\}\notag\\
    +\frac{1}{2\,\dy^2} &\{D_u(t,x,y)[u(t,x,y-\dy) - 2u(t,x,y) + u(t,x,y+\dy)] \notag\\
    &+D_u(t,x,y-\dy)[u(t,x,y-\dy) - u(t,x,y)]\notag\\
    &+D_u(t,x,y+\dy)[u(t,x,y+\dy) - u(t,x,y)]\}.
\end{align}
Here, $D_u$ can depend on space (explicitly or implicitly as a function of the variables $\vec{u}$), though the approximation reduces to a standard second-order central difference approximation if $D_u$ is independent of space. At boundaries, ghost nodes are introduced to allow for the enforcement of boundary conditions through modification of the finite difference schemes. VisualPDE accommodates periodic, Neumann and Robin boundary conditions in this way. Dirichlet boundary conditions are also supported, though these do not modify the finite difference operators and are instead implemented by fixing the values at these nodes. Boundary conditions can be fully customised by the user via the VisualPDE interface, including mixed conditions on different interfaces, and arbitrary inhomogeneous boundary terms (e.g.~time dependent boundary conditions are obtained just by writing some function of $t$ when specifying the boundary condition).

Domains that are not rectangular are implemented by casting them as subsets of a rectangular domain, with individual points in the finite difference discretisation included or excluded via an indicator function. As this enables users to specify general domains that need not have smooth boundaries, boundary conditions on non-square domains involving derivatives should be interpreted with due care.

\subsubsection{Timestepping}
With space discretised as above, the resulting system of ordinary differential equations are solved by discretising in time and employing one of four explicit, fixed-timestep finite difference schemes: forward Euler, two-step Adams--Bashforth, the midpoint method (also called the modified Euler method) and the four-step Runge--Kutta method (often known as RK4). These schemes are each associated with benefits and drawbacks, with RK4 conferring the broadest stability, the highest order accuracy and the greatest computational cost, while forward Euler represents the computationally simplest option but scales least favourably with the timestep $\dt$.

Forward Euler is the default option for the majority of the examples in VisualPDE due to its simplicity and somewhat surprising reliability in practice. For completeness, this scheme approximates
\begin{equation}
    \left.\pd{u}{t}\right\rvert_{(t,x,y)} \approx \frac{u(t+\dt,x,y) - u(t,x,y)}{\dt},
\end{equation} 
which is first-order accurate in $\dt$. At any point a user can select from any of the timestepping schemes available, including midway through a simulation, which might be needed to satisfy a user's requirements of accuracy, stability or curiosity. In this way, VisualPDE enables the simple, accessible exploration of the properties of numerical schemes in practice.

\rev{\subsubsection{Validation}
We have validated the results from VisualPDE through a number of comparisons to published results, as well as our own finite difference, spectral, and finite element implementations. We have also included a `Numerical Methods' section on the website, which we plan to expand with further examples discussing aspects of the methods used. Currently this section consists of an \hereVPDE[example]{https://visualpde.com/numerical-methods/validating-VisualPDE.html} that explores three equations with analytical solutions that can be quantitatively compared to numerical solutions. Importantly, this also invites the reader to explore changing the timestep or timestepping scheme, in particular in the context of the Schr\"{o}dinger equation where a total mass of the wavefunction is only approximately conserved.}

\subsection{WebGL implementation \& technical considerations}
Solving the large system of ordinary differential equations (ODEs) that arises from the spatial discretisation and displaying their solution presents non-trivial computational challenges. This is compounded by the desire for VisualPDE to run in browsers across a broad range of devices, including mobile phones, with a high level of interactivity and speed. To overcome this challenge, VisualPDE exploits the significant and widespread capabilities of graphics processing units (GPUs), present in some form on essentially all modern computing devices. 

In essence, VisualPDE achieves this by casting the problem of timestepping ODEs as one of iterated image manipulation. Floating-point image textures are used to represent the computational domain, whereby each pixel corresponds to a point in the spatial discretisation. Taking advantage of the massively distributed parallel architecture of modern GPUs, the points in the spatial domain are updated in parallel every timestep. The speed of this image-based approach typically entails that VisualPDE can advance the solution a great many times each second, with recent devices (including mobile phones) being capable of upwards of 24,000 timesteps per second in representative simulations. This amounts to performing hundreds of timesteps every time the user's device requests an update to the solution being displayed on the screen, which typically occurs 30--60 times each second. We refer directly to the GitHub code \citep{github} for further details on the implementation, which makes use of the Three.js library for interfacing with WebGL.

The parallelism that drives the responsiveness and speed of VisualPDE does, however, come with its own limitations. One significant limitation is that the independent updating of each pixel prevents the simple, efficient implementation of implicit timestepping schemes, which typically update every point at once as the solution of a coupled system of equations. This is the reason why VisualPDE only makes use of explicit timestepping schemes, though we note that the conditional numerical stability associated with these schemes is largely mitigated by the frequency at which timesteps can be taken, with small $\dt$ thereby not detracting from the user experience.

Another barrier posed by the use of GPUs is the lack of widespread support for double-precision floating point textures. While these are available on some devices, they do not appear to be ubiquitously supported. Thus, VisualPDE makes exclusive use of single-precision arithmetic to ensure a consistent experience across devices, though we expect to transition to double-precision arithmetic as support widens. Simulations using the website do remarkably well in quantitatively matching more accurate numerical methods using standard double-precision methods, though we would advise some caution if an example demands particularly high accuracy and precision.

Providing a consistent experience for all users has presented a number of additional challenges, not least of which are vendor-specific differences in the capabilities and behaviours of various devices. For instance, many mobile devices do not automatically interpolate low-resolution textures onto the display, while many laptop and desktop computers do. To circumvent this, VisualPDE attempts to detect your device and its capabilities and implements bespoke interpolation where necessary. Despite the varied idiosyncrasies of popular devices, extensive testing suggests that the functionality of VisualPDE is maintained across many kinds of devices, with the curated examples configured to provide a smooth, interactive experience on even relatively low-end hardware.

\section{Interacting with VisualPDE}\label{sec:interaction}

\subsection{Design of the user interface}
The VisualPDE user interface (UI) consists of two halves: a static website, with colourful tutorials, examples and user guides, and the interactive simulation page, where PDEs are solved in real time inside the browser window. Our guiding philosophy for both is to let intuition guide the user: to be minimal yet feature rich.

The website is written using the static site generator Jekyll \citep{jekyll}, which compiles HTML files from simple text-based markdown scripts. Combined with MathJax \citep{cervone2012mathjax} to enable \LaTeX-style mathematical typesetting, it allows pages to be written by authors without technical HTML or CSS knowledge, scaling gracefully to different device sizes.

\begin{figure}
    \centering
    \includegraphics[width=0.48\textwidth]{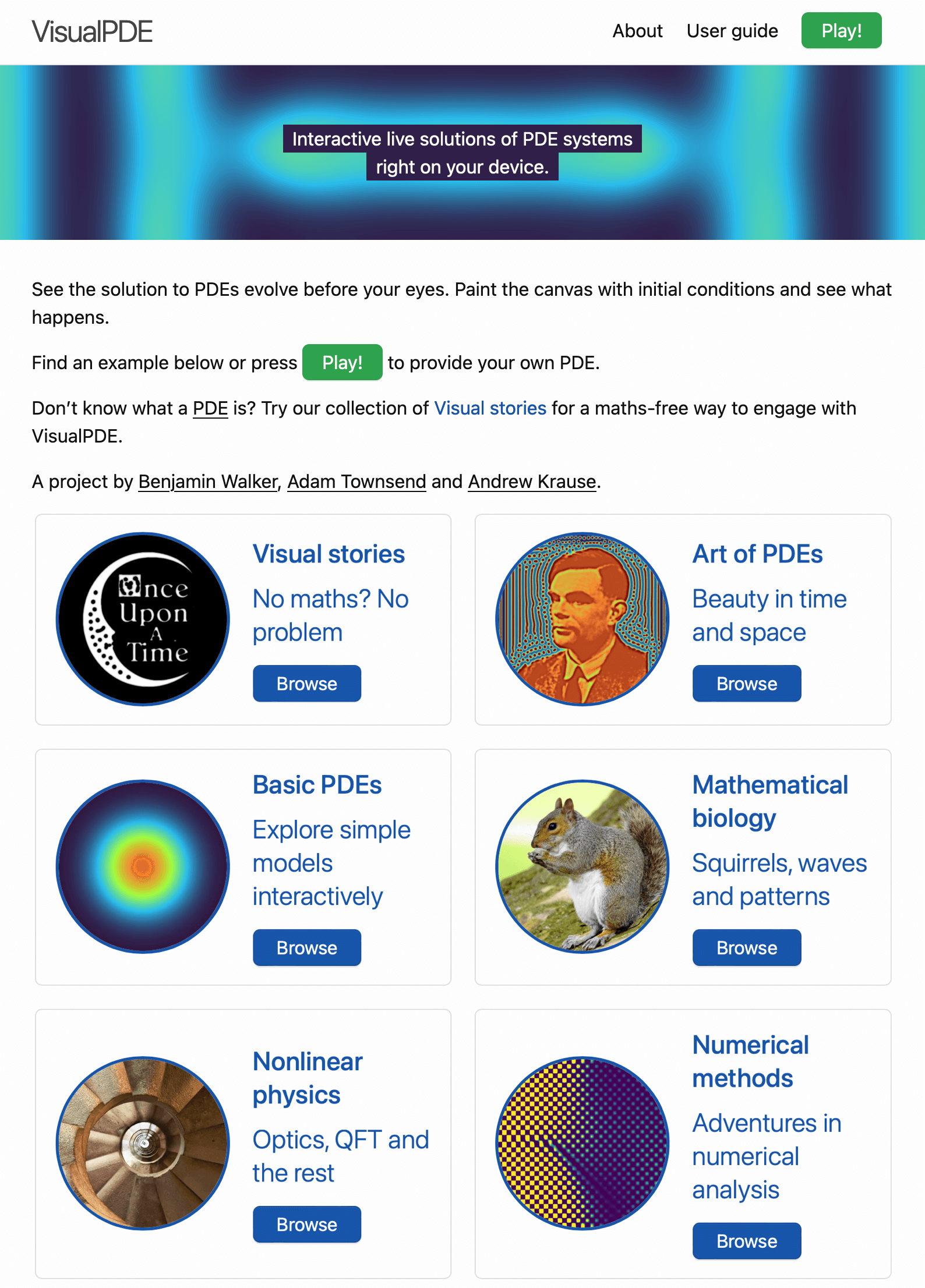}\qquad
    \includegraphics[width=0.48\textwidth]{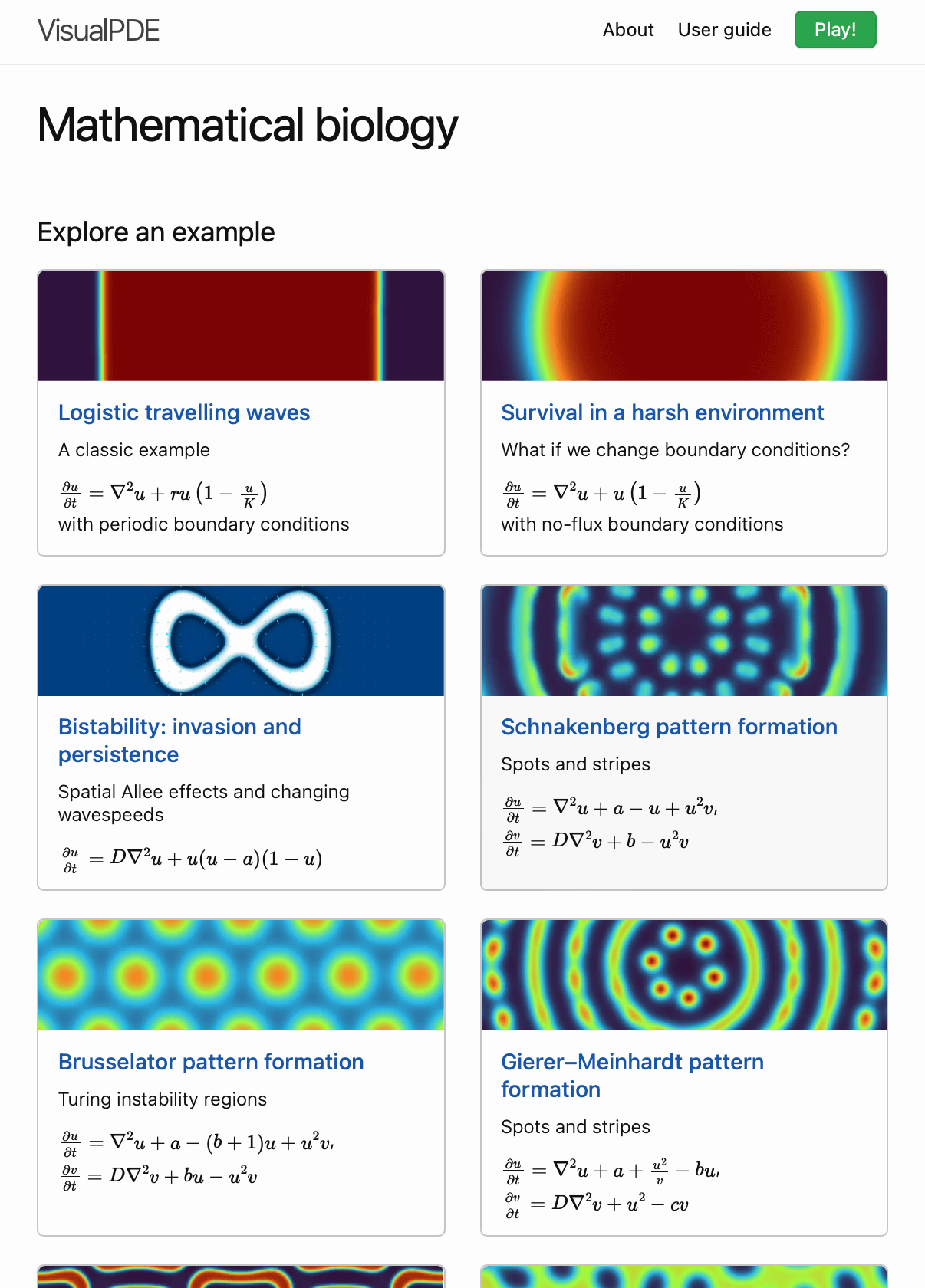}
    \caption{The VisualPDE static website. \textbf{Left:} The homepage presents \rev{several} collections of \rev{examples organised by theme}. \textbf{Right:} The mathematical biology tutorial collection starts with simple models and becomes increasingly complex. Attractive simulation screenshots indicate the behaviour the user might observe. Captured July 2023.}
    \label{fig:collection}
\end{figure}

\begin{figure}
    \centering
    (a) \includegraphics[width=0.46\textwidth,valign=c]{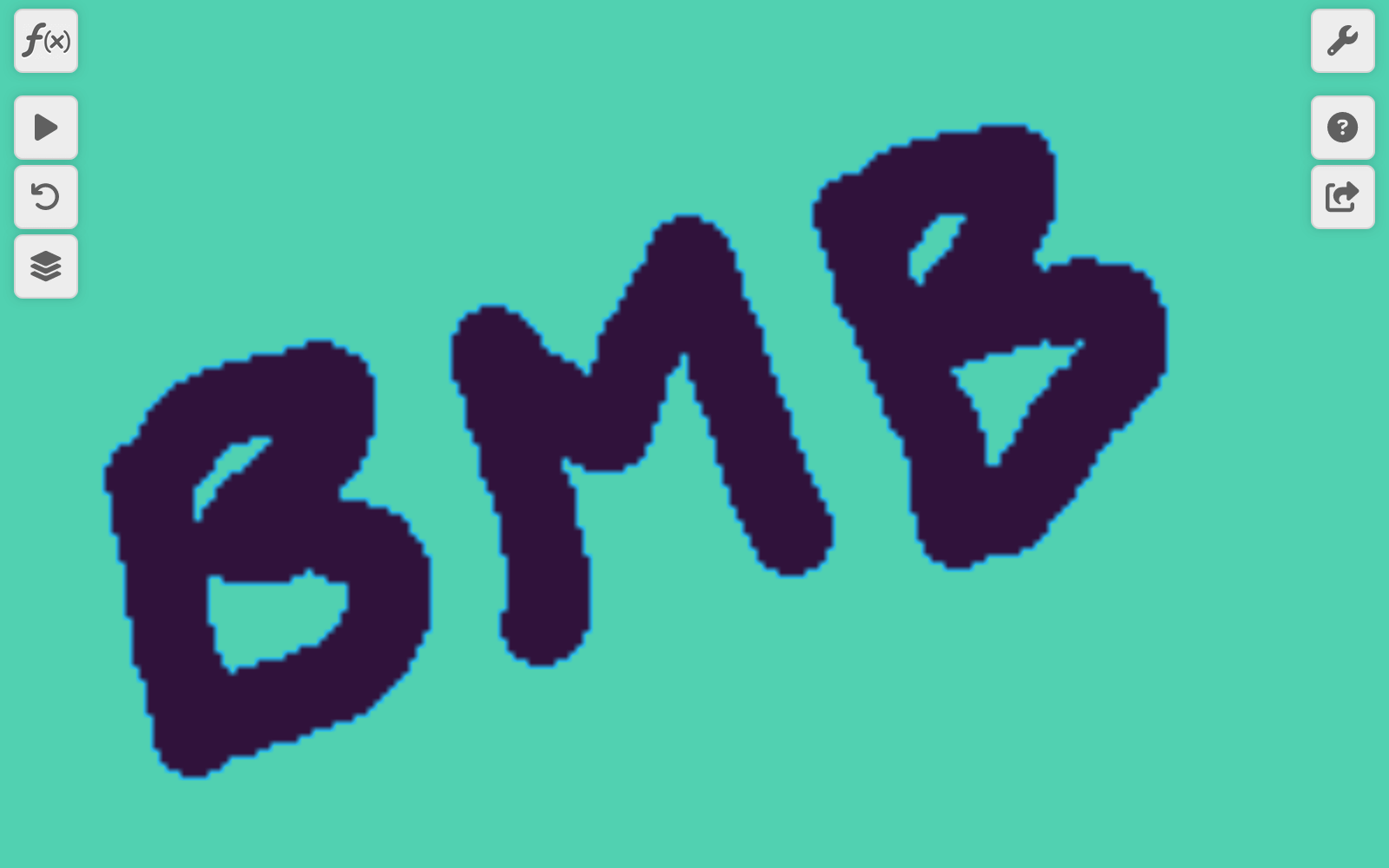}\quad
    (b) \includegraphics[width=0.46\textwidth,valign=c]{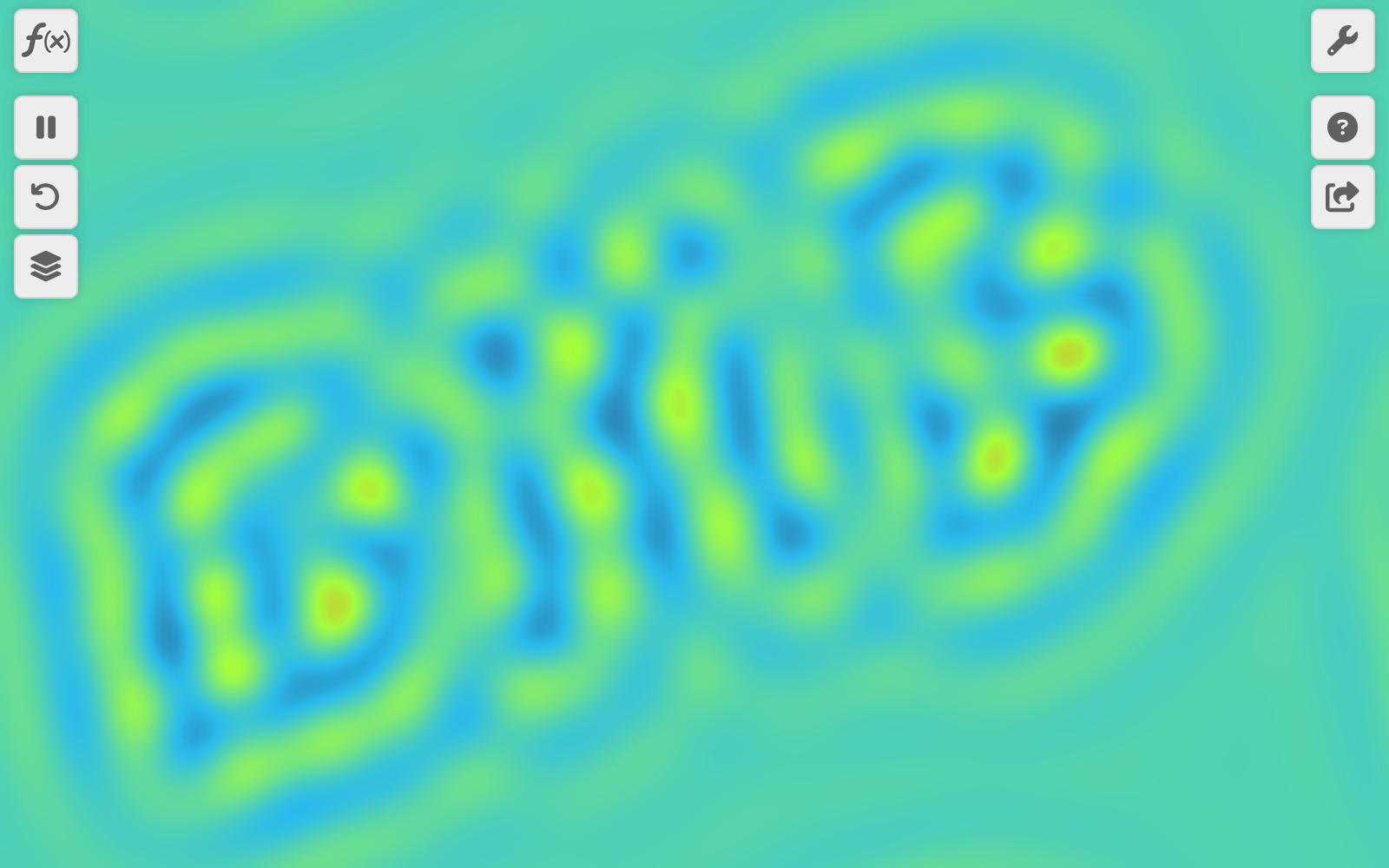}    
    
    \vspace{5mm}
    (c) \includegraphics[width=0.46\textwidth,valign=c]{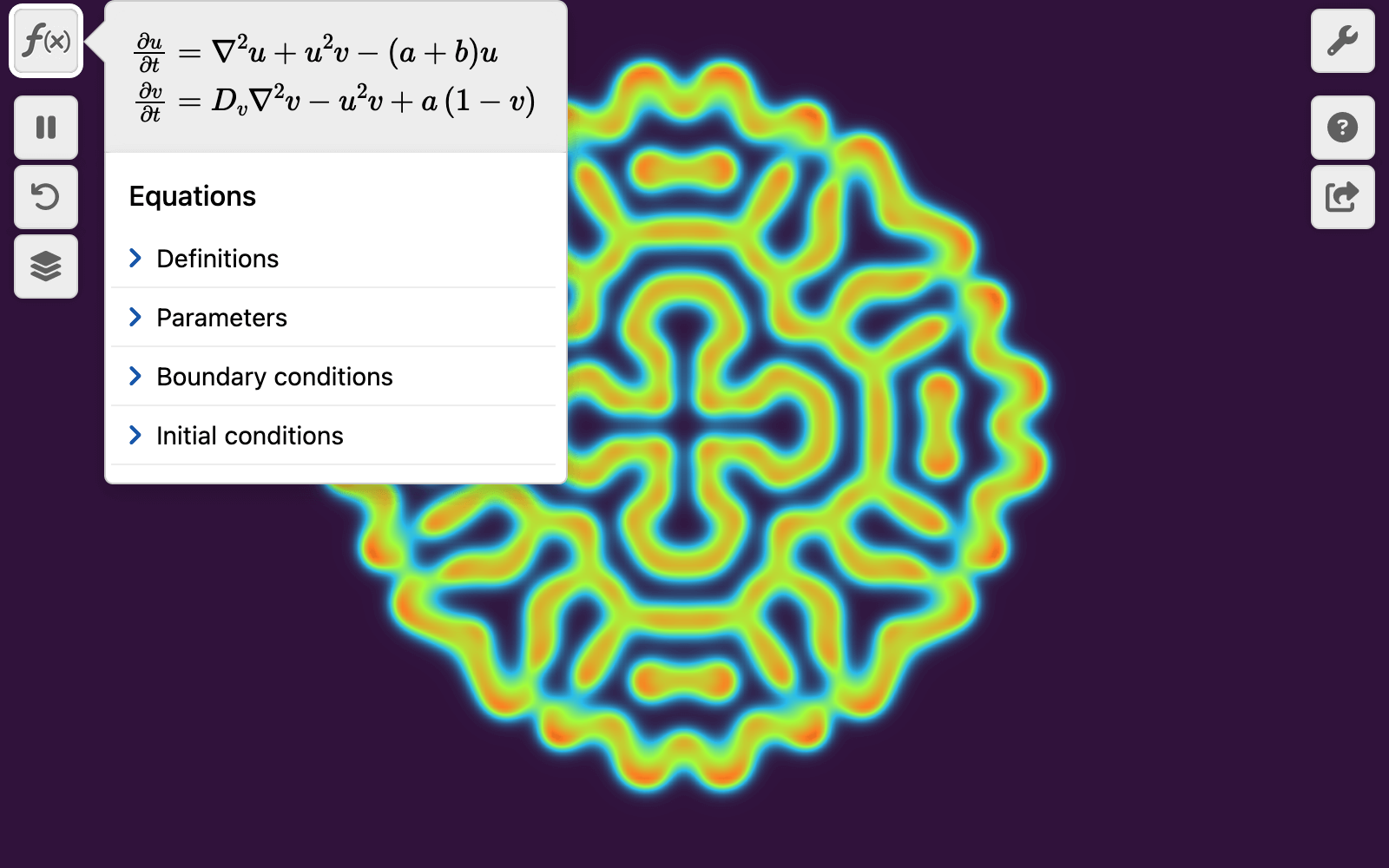}\quad
    (d) \includegraphics[width=0.46\textwidth,valign=c]{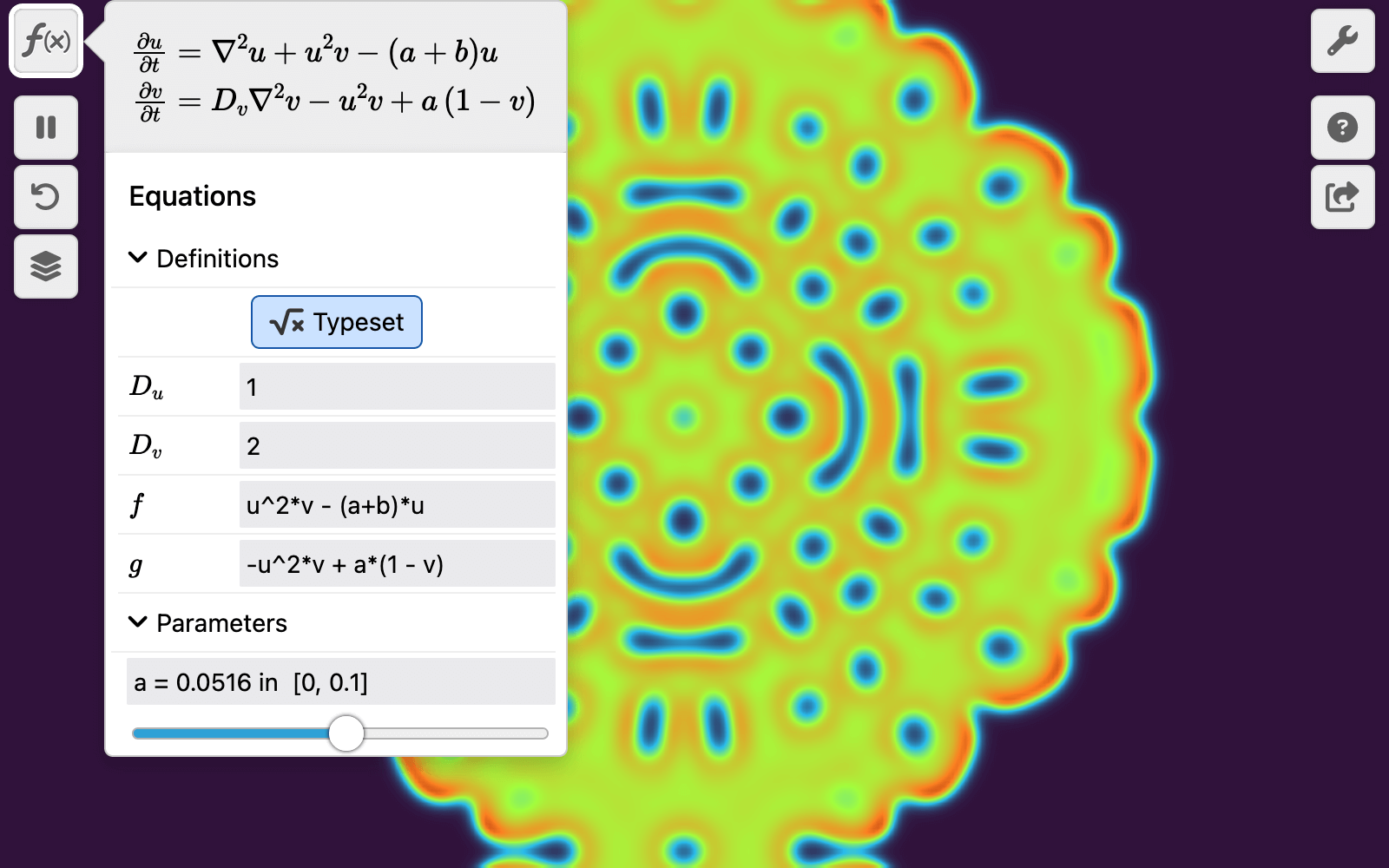}

    \vspace{5mm}
    (e) \includegraphics[width=0.46\textwidth,valign=c]{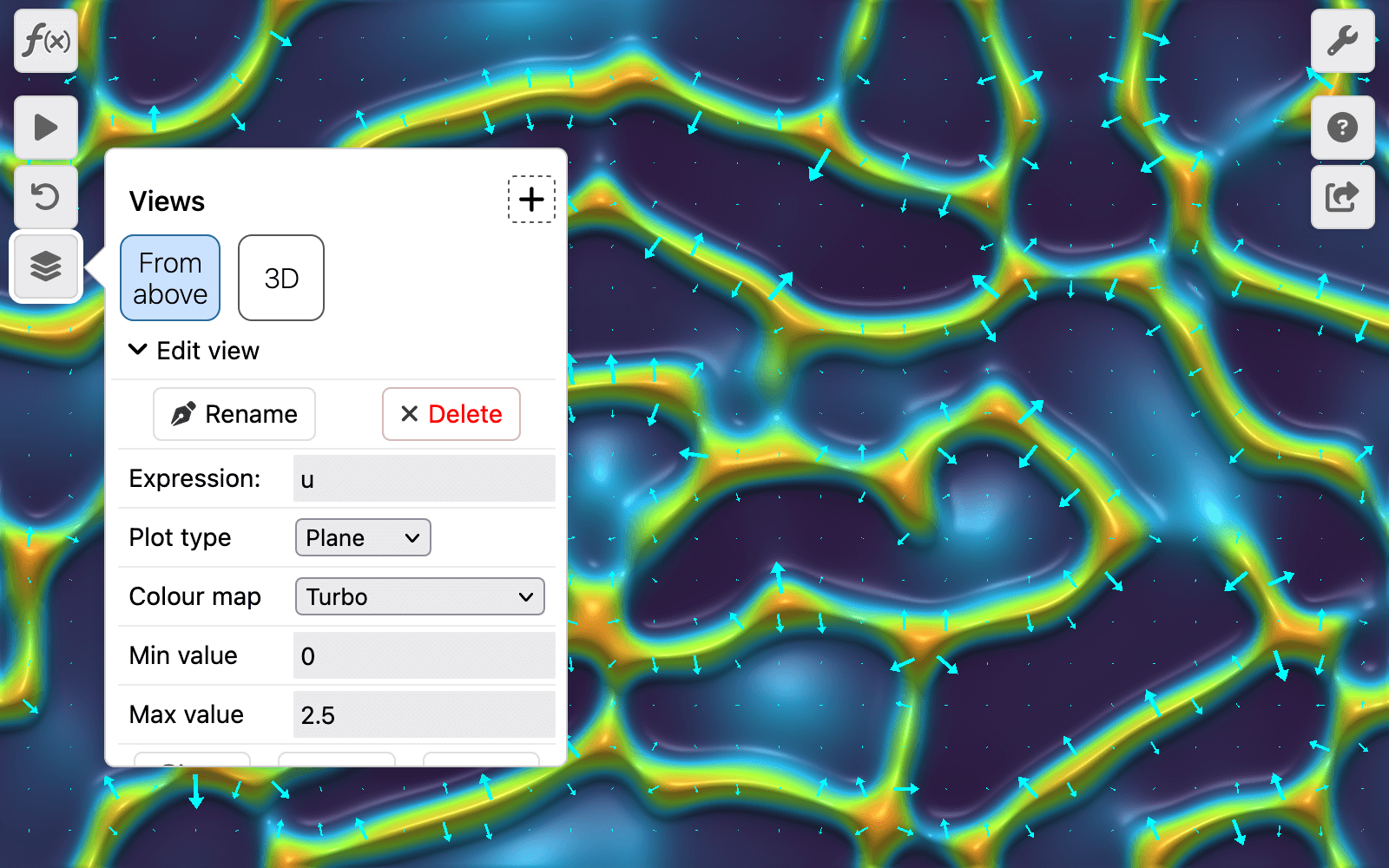}\quad
    (f) \includegraphics[width=0.46\textwidth,valign=c]{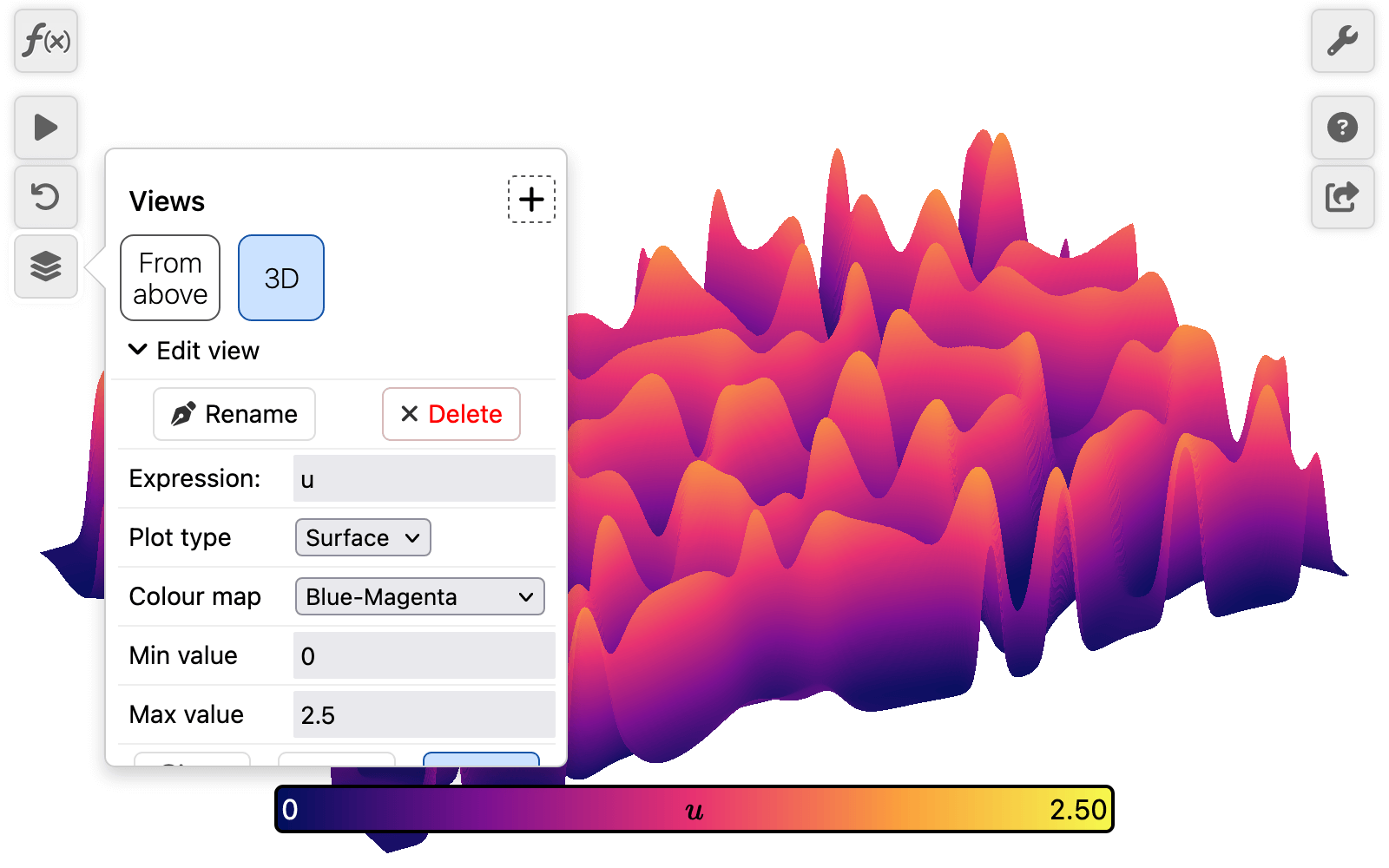}

    \vspace{5mm}
    (g) \includegraphics[width=0.46\textwidth,valign=c]{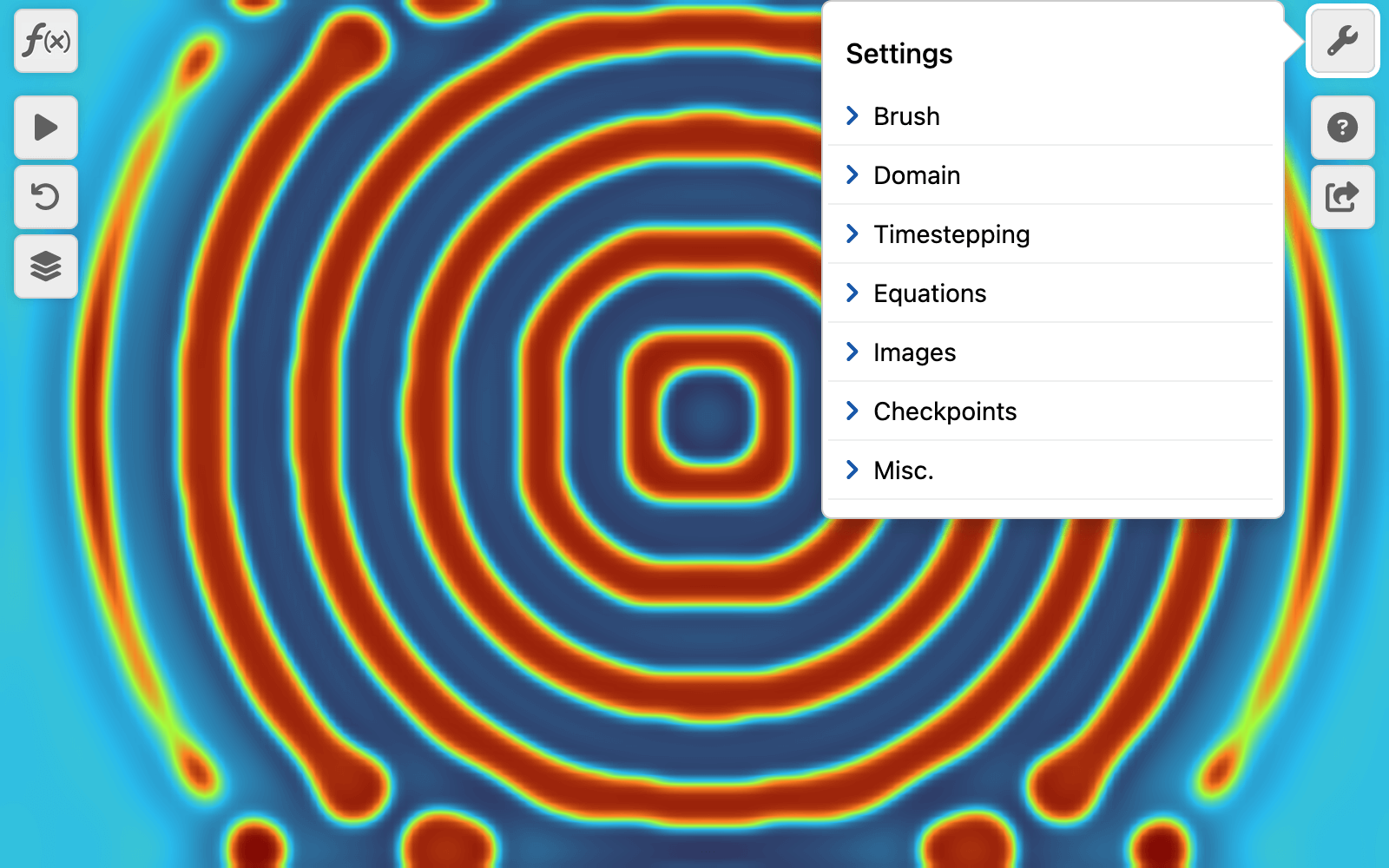}\quad
    (h) \includegraphics[width=0.46\textwidth,valign=c]{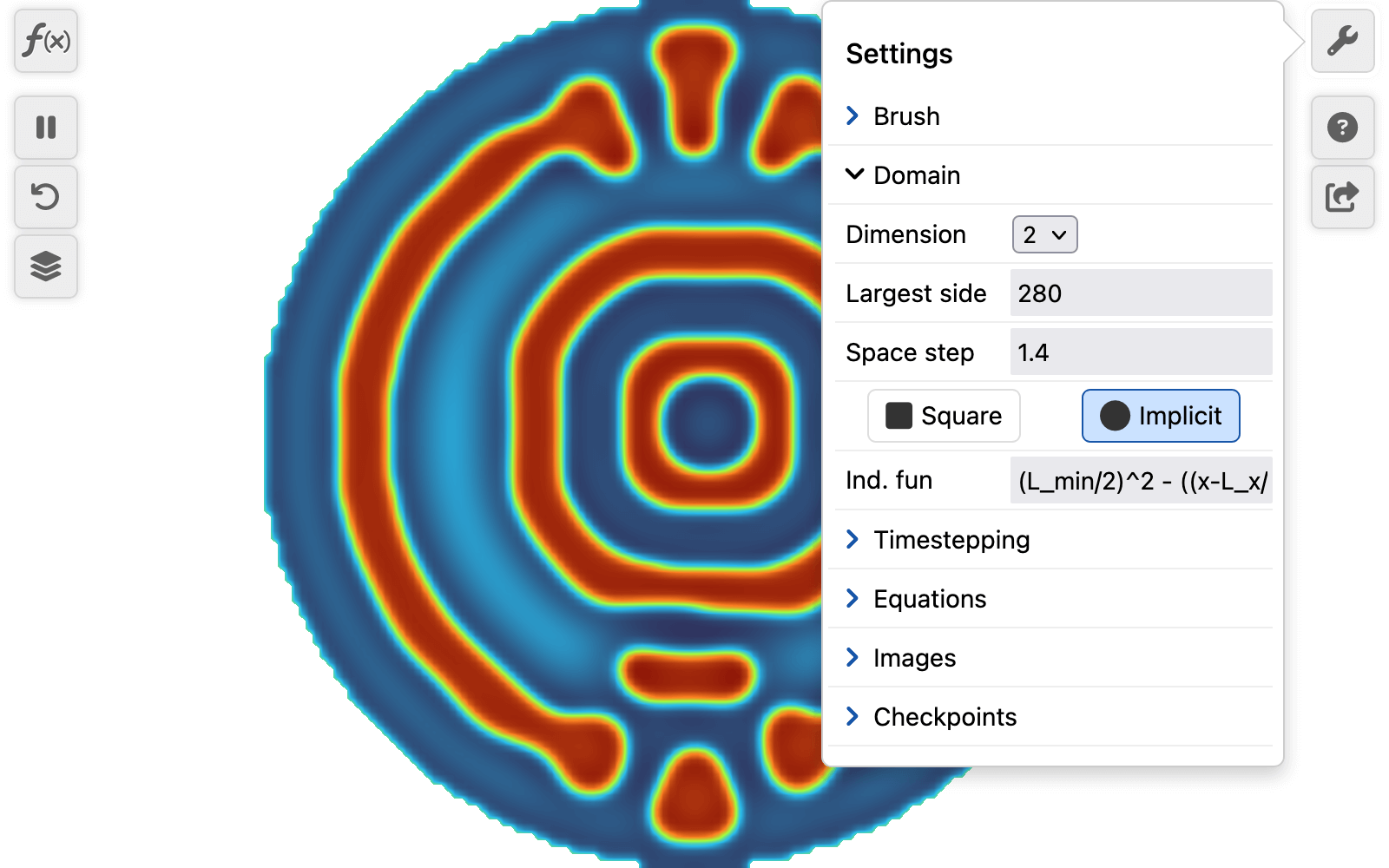}
    
    \caption{The VisualPDE simulation page. \textbf{(a)} The user can pause the simulation and paint an initial condition on the screen using the cursor or their finger. \textbf{(b)} By clicking `play', the simulation starts (here, the Brusselator). \textbf{(c)} The equations panel allows users to edit the equations and parameters being simulated. \textbf{(d)} Editing the definitions and parameters in real time allows the user to see the effect immediately (here, in the Gray--Scott model, $a$ has been increased.) \textbf{(e)} The Views panel allows for selection between (potentially preset) quantities to visualise, as well as providing a selection of colour maps, scales, lighting and other options. Here, we are simulating Keller--Segel chemotaxis with a 3D lighting effect and an overlaid vector field $(-\partial u/\partial x, -\partial u/\partial y)$. Some options in the panel require scrolling down to see. \textbf{(f)} The same solution at the same time as (e) but viewed as a surface plot, with a different colour map and with a colour bar. \textbf{(g)} The settings panel allows for changes in domain shape, timestepping scheme, brush sizes when painting with the cursor/the user's finger, and any images used as input. Here we see FitzHugh--Nagumo on a rectangular domain. \textbf{(h)} The same system as (g) but on a circular domain. Captured July 2023 on a $800\times500$px screen.}
    \label{fig:sims}
\end{figure}

The front page of the website presents links to a number of collections that each contain guided tutorials on a particular PDE system (see \cref{fig:collection}) or several related models. These collections start with basic linear PDEs, followed by more complicated systems in mathematical biology,  nonlinear physics, and numerical analysis. The curated examples are largely chosen to demonstrate different phenomena that we feel benefit from being explored through interactive visualisations, as well as to illustrate different features of VisualPDE. We also include our innovative curiosity-driven `Visual Stories', discussed in \cref{sec:examples}. Importantly, these collections of examples are not meant to be exhaustive, and features are provided to make crafting and sharing entirely new models seamless and simple.
Each tutorial contains links to the simulation page, with preset equations, parameters, and other settings. The user can then change these on the simulation page and see how the system solution changes. \rev{Importantly, every example can be transformed into every other example on the website simply by clicking and typing within the interface. This includes changing the colour scheme and other visualisation options under the Views menu at the bottom of the left column of buttons (see \cref{fig:sims}e).}

The simulation page considers the entire device browser window as the domain on which to solve the PDE. Optimised for both mobile and static devices, the initial user experience is a blank coloured page with icons on the left and right. An invitation to click or tap (device depending) prompts the user to paint the screen using the cursor or their finger to set a forcing or initial condition of the system (\cref{fig:sims}a). Painting can be done at any time, live, with a configurable brush; using the pause and play icons allows for more complex painting that evolves live when the simulation is resumed (\cref{fig:sims}b).

Although the static website links to many preset simulations, the equations are entirely configurable from within the simulation page. The equations panel (\cref{fig:sims}c--d) allows for live alteration of all the equations within the form discussed in \cref{sec:which-pdes}. Expressions are instantly typeset in MathJax both for familiarity and to assure the user that VisualPDE has interpreted their expression correctly. As a user modifies a part of the equations, the relevant typeset portion of the equation is highlighted to guide the user. Sliders for parameters can be created easily by the user, similar to those seen in popular online 2D plotting software, allowing for interactive, intuitive parameter sweeps. The limits of these sliders can be fully customised so that, for example, educators can suggest ranges for students to look within. Boundary conditions for each species can be individually set to periodic, Dirichlet, Neumann, Robin or a specified combination through a drop-down menu. With these features, the UI is intended to make explorations of the impact of boundary conditions, or the presence of bifurcations, intuitive and simple to play with.

VisualPDE presents many options for visualising solutions. The Views panel (\cref{fig:sims}e--f) allows users to specify which expression they want to see mapped onto the domain: this can be customised and can include any nonlinear combination of species or even explicit time or space dependence. Although the default for 2D systems is to display the solution as a 2D image, the solution can also be visualised in 3D on a surface, or for 1D models, on a line. An adaptive colour bar can be displayed and customised, with a wide variety of colour maps available to cater for user preference and enhance accessibility \citep{smith2015better}. For even richer visuals, contours and 3D-effect lighting can be added, along with custom, solution-dependent vector fields and graphical overlays. All these viewing options can be saved as preset views and given a custom name (in \cref{fig:sims}e--f, `From above' and `3D'), so that simulations can be shared and understood easily.

\begin{figure}
    \centering
    \includegraphics[width=0.33\textwidth,valign=c]{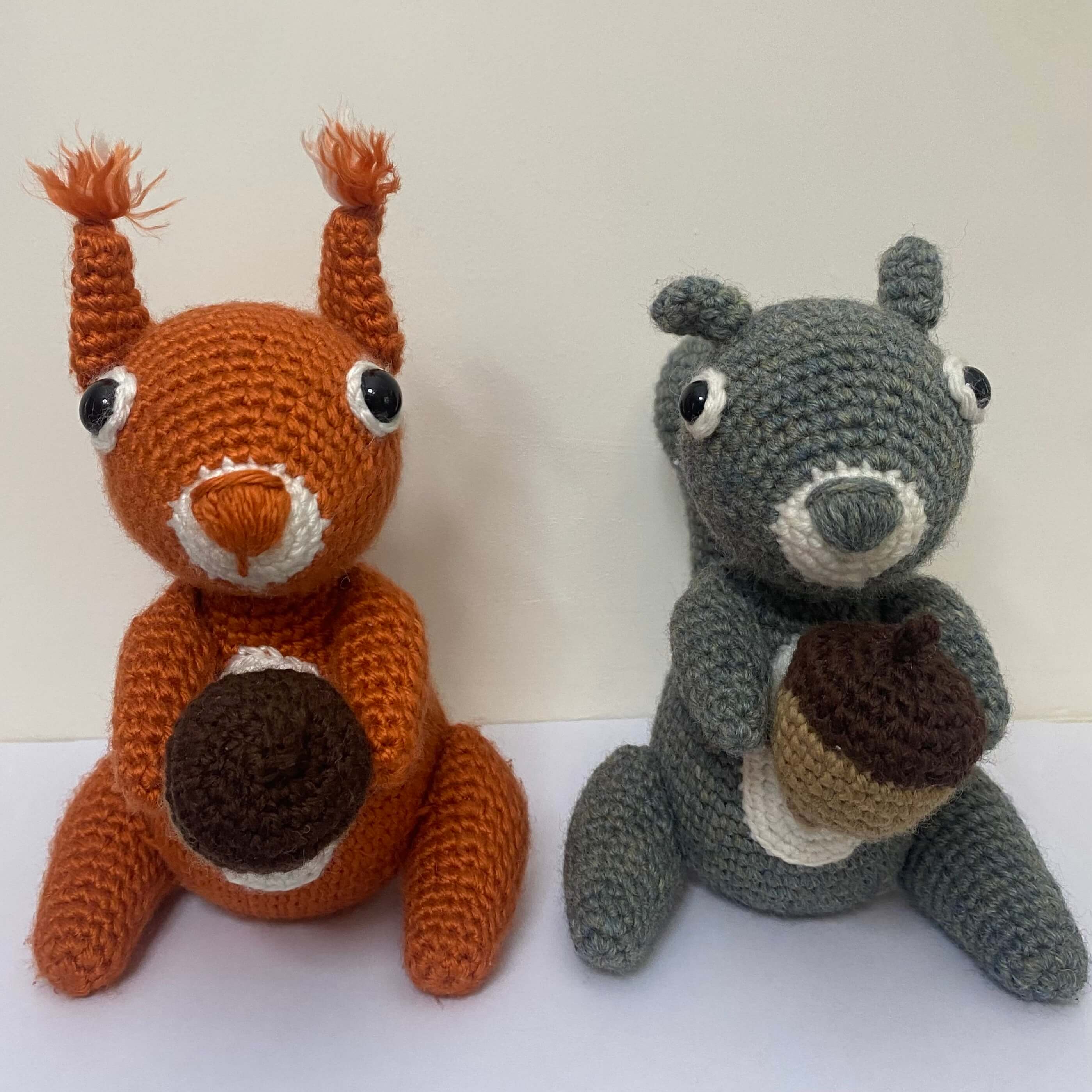}
    \; $\to$ \includegraphics[width=0.26\textwidth,valign=c]{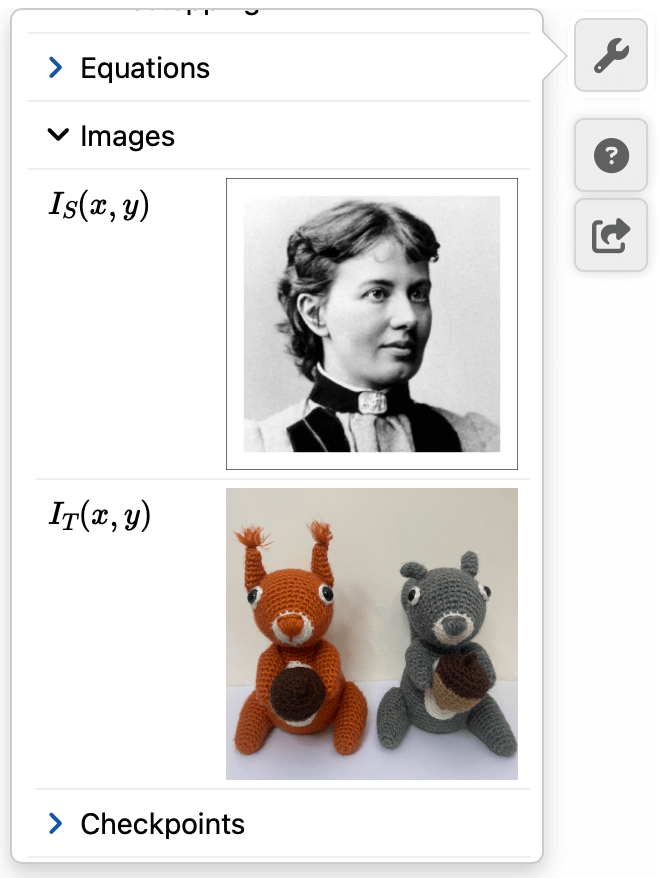}   
    $\to$ \; \includegraphics[width=0.33\textwidth,valign=c]{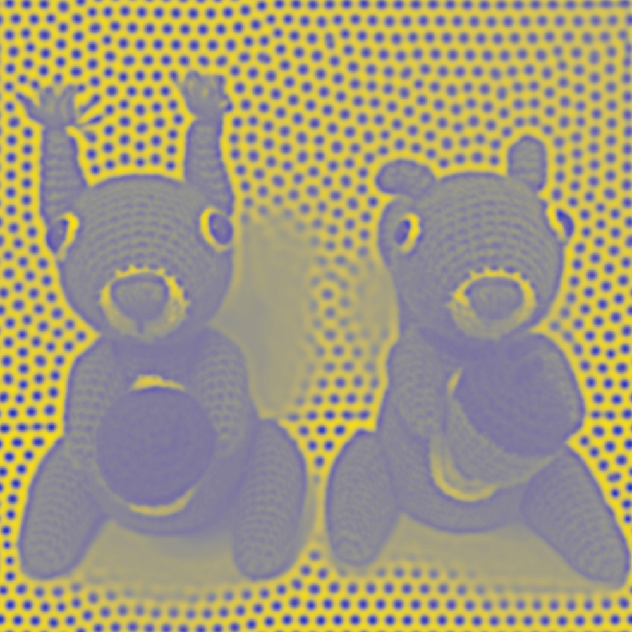}  
    \caption{Turingify your friends: Users can take photos on their phone and use them as forcing in their PDEs. \rev{Here, the \hereVPDEfig[Turing on Turing]{https://visualpde.com/art-pdes/turing-face.html} example on VisualPDE was customised with happily coexisting red and grey squirrels}.  \rev{The user can also select a predefined image such as Sofya Kovalevskaya or Alan Turing (not shown).} Original image of squirrels courtesy of Elizabeth Brocklebank.}
    \label{fig:squirrels}
\end{figure}
\footnotetext{\url{https://visualpde.com/art-pdes/turing-face.html}}

Finally, the settings menu (\cref{fig:sims}e) contains slightly more advanced options. Numerical options include the dimension, step size and shape of the domain, as well as the timestepping scheme and timestep size. Presentation options include custom choices of letters for the species functions (by default $u$, $v$, $w$ and $q$) to better match any source material; the shape and size of the brush for painting; labels showing the elapsed time and integral of the displayed solution. Checkpoints can be set so the user can restart the simulation to a given timestep, rather than to the beginning of the simulation. A particularly fun option here is to upload a photograph to use as a spatial function inside an equation (\cref{fig:squirrels}). By default, Sofya Kovalevskaya and Alan Turing are functions $I_S(x,y)$ and $I_T(x,y)$, but clicking or tapping on their photos prompts the user to either upload a different photo (on a computer) or to take a photo using their camera (on a phone). We consider this a fun hook into the software.

User feedback regarding error handling proves a particular challenge in presenting a user-friendly interface, especially for users who may be unfamiliar with the underlying numerical methods. VisualPDE provides custom messages for syntax errors, and for solutions blowing up. Numerical solutions (\cref{sec:numerical-methods}) can be fragile, in strong contrast to 2D plotting software that novice users may bring their intuition from. Ideally users should be told whether an infinite solution has been reached because of numerical instability or because the solution is truly exponential growth. This is beyond the scope of VisualPDE at the moment, but the omnipresent help icon and the error messages both lead to discussions on numerics that may be helpful, including practical tips for improving numerical stability. Preset simulations where parameters are constrained by sliders and where equations are uneditable (as in the Visual Stories) may therefore be more beneficial to novice users.

\subsection{Sharing VisualPDE models}

\begin{figure}
    \centering
    (a) \includegraphics[width=0.43\textwidth,valign=c]{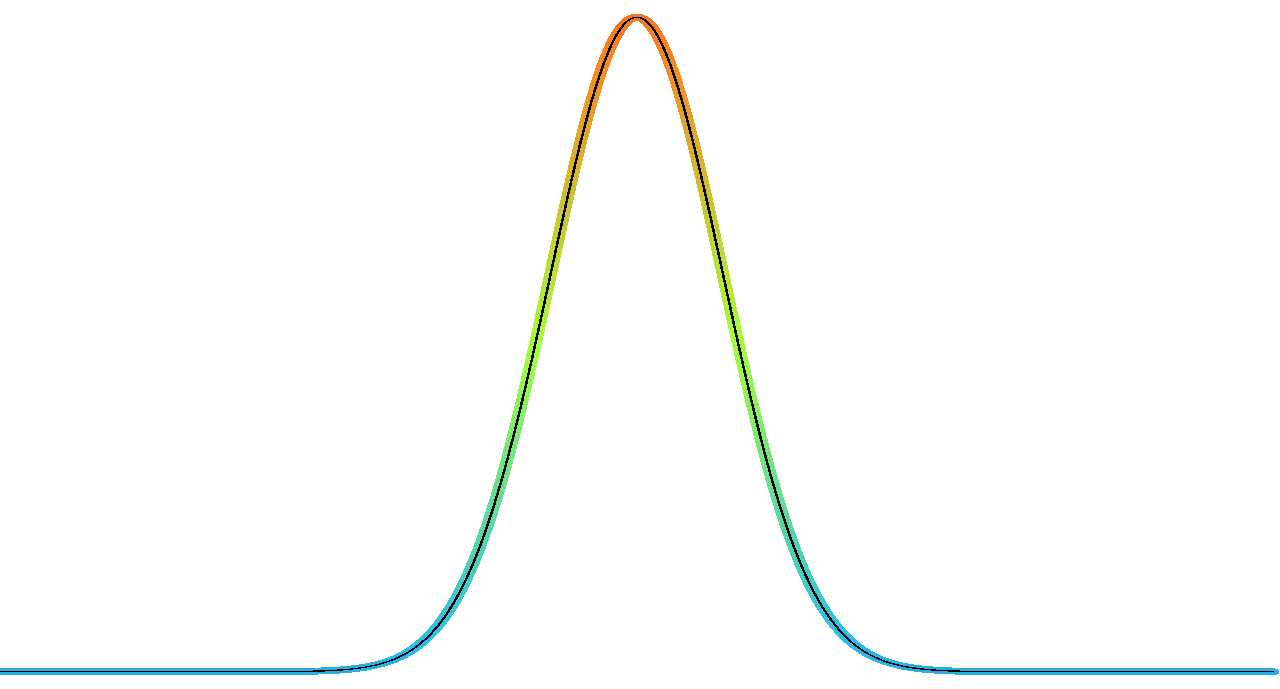}\, $\to$ \,
    (b) \includegraphics[width=0.43\textwidth,valign=c]{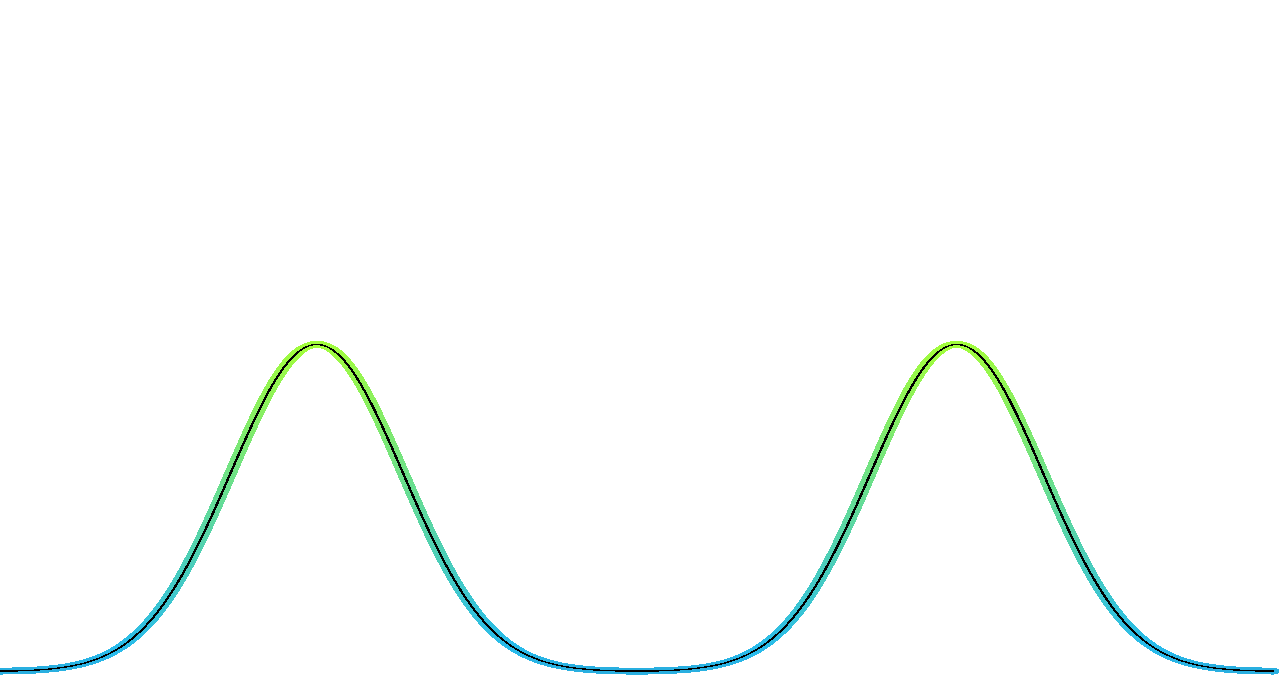}    \\
    \vspace{5mm}
    (c) \includegraphics[width=0.43\textwidth,valign=c]{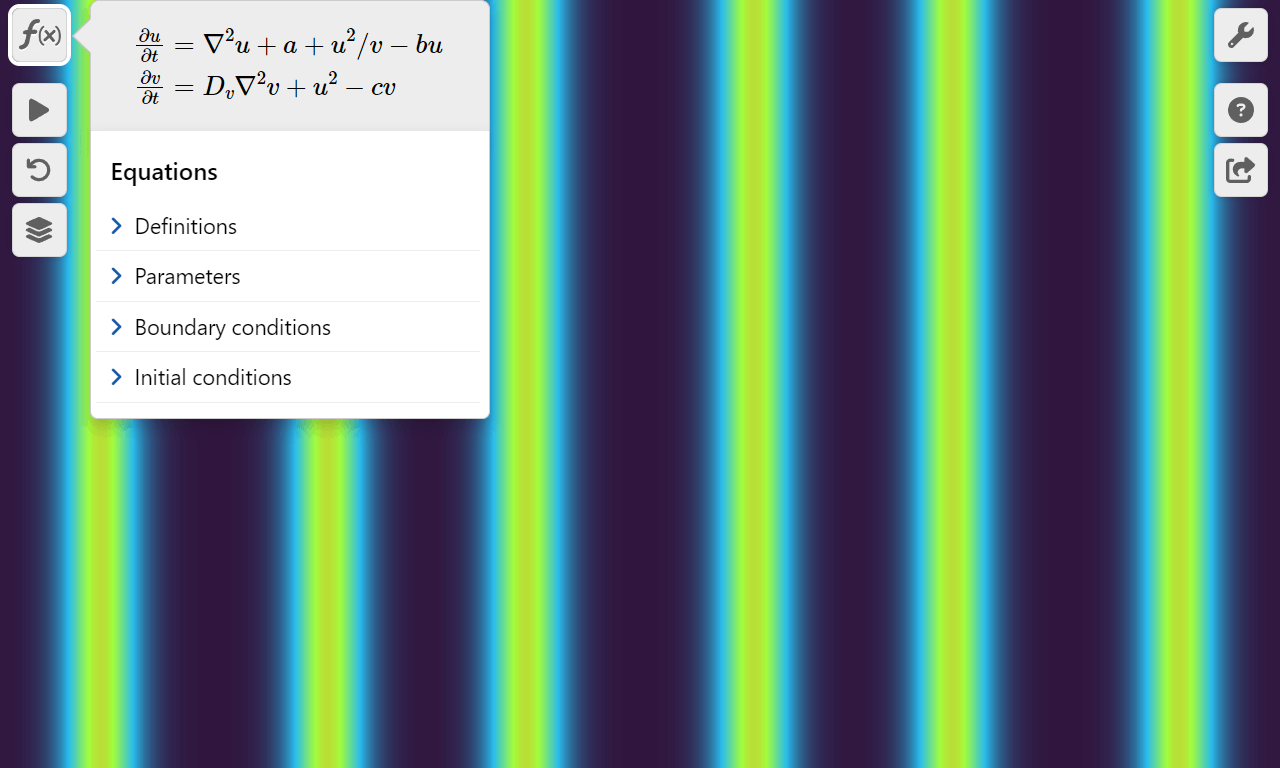}\, $\to$ \,
    (d) \includegraphics[width=0.43\textwidth,valign=c]{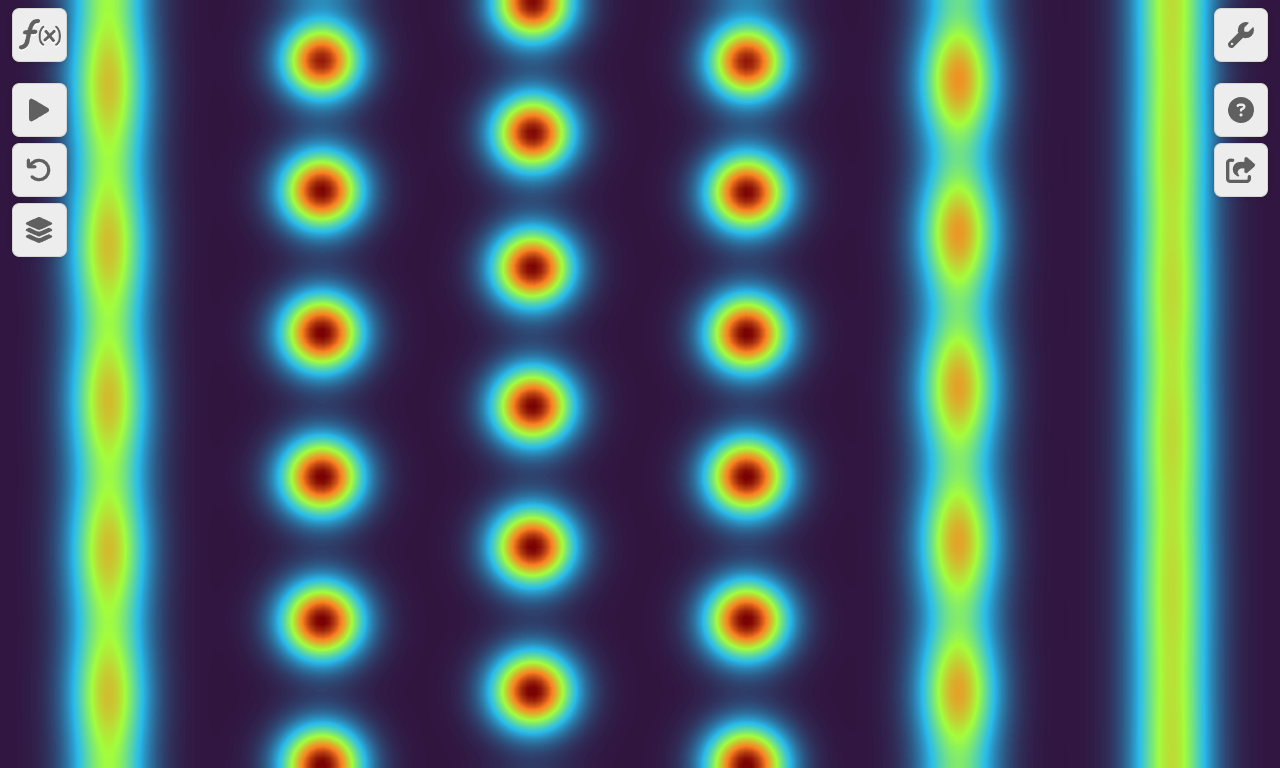}\\    
    \vspace{5mm}
    (e) \includegraphics[width=0.43\textwidth,valign=c]{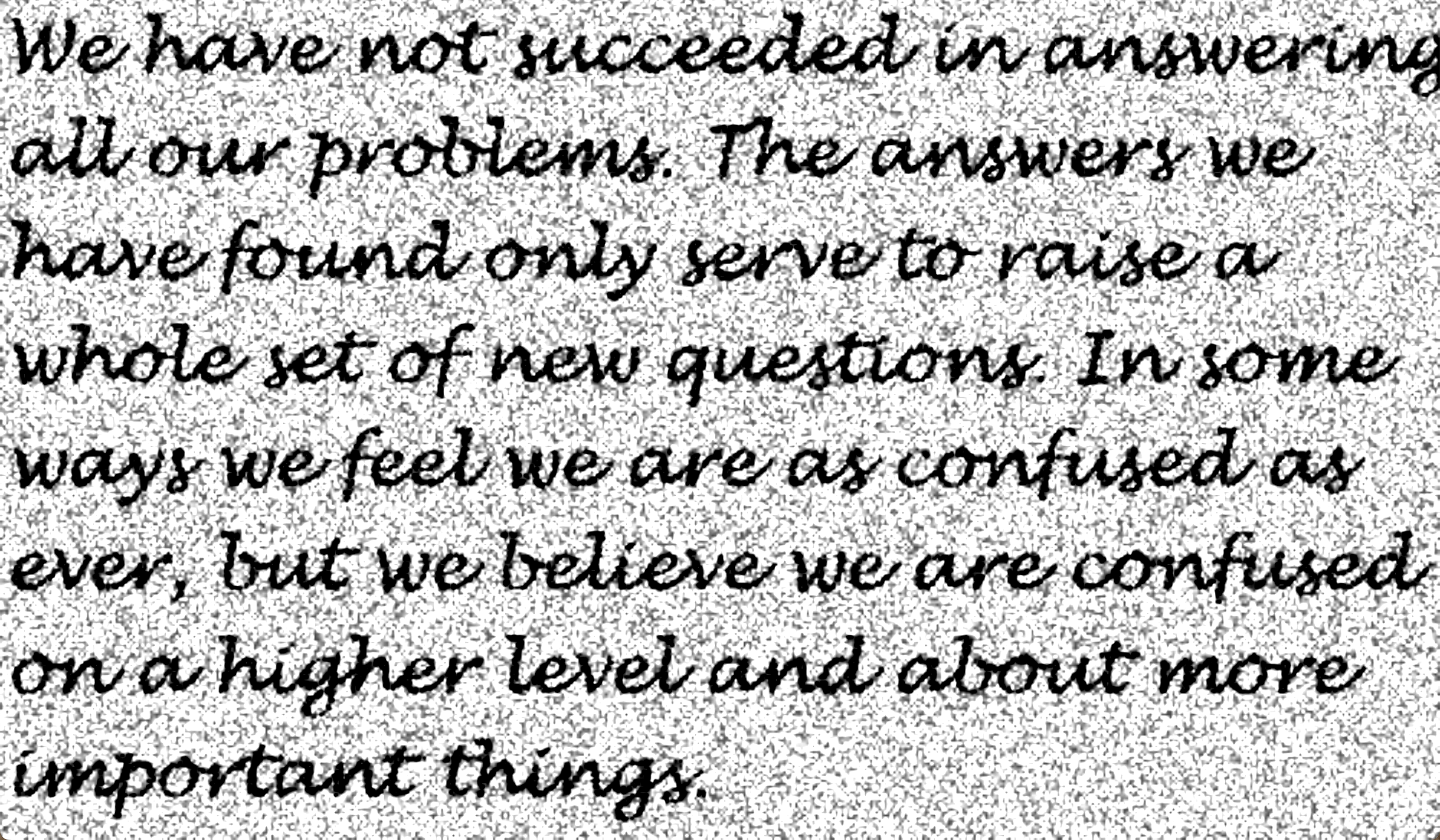}\, $\to$ \,
    (f) \includegraphics[width=0.43\textwidth,valign=c]    {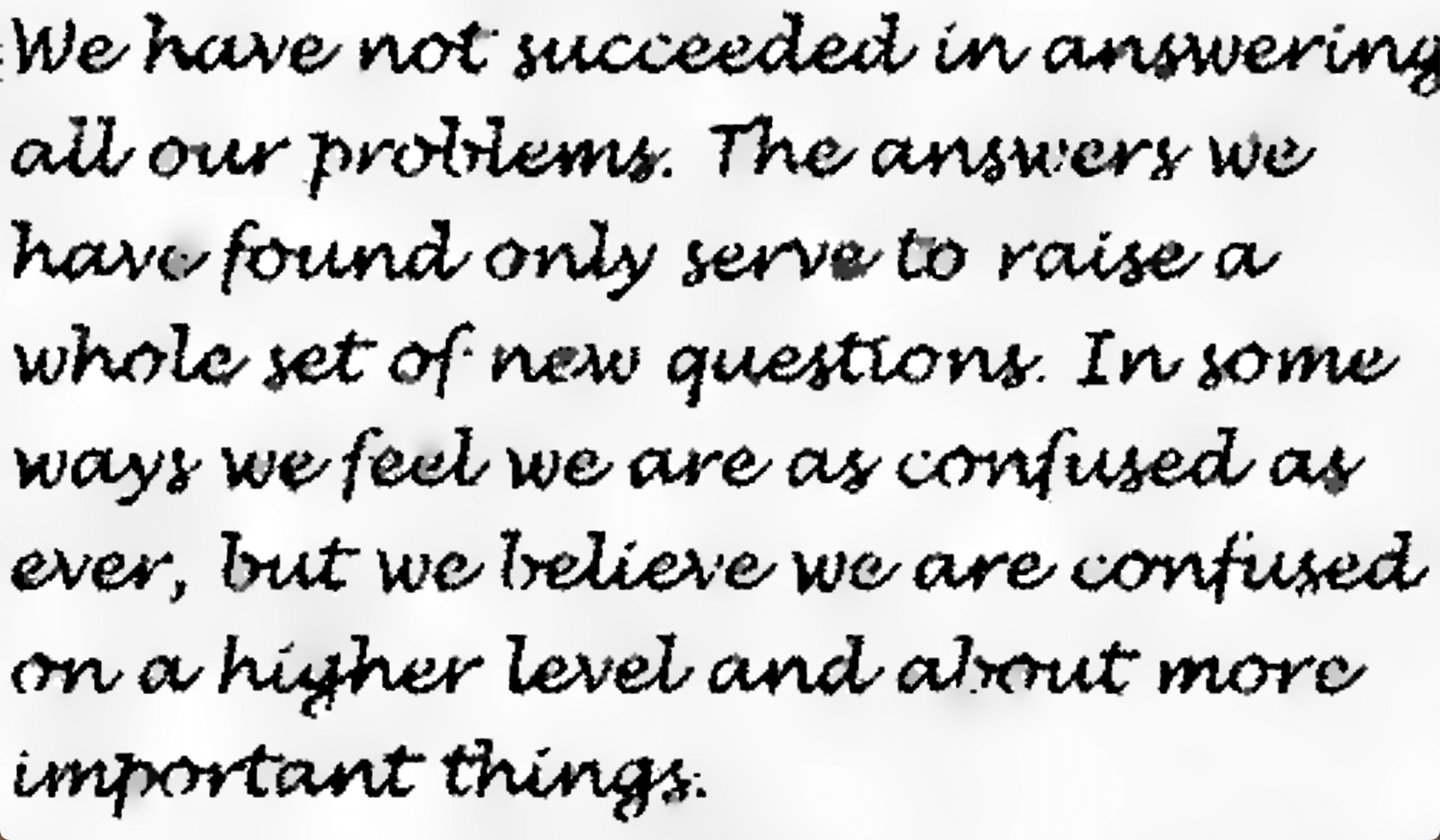}    
    \caption{Example simulations from teaching. \textbf{(a)--(b)} An \hereVPDEfig[example d'Alembert solution]{https://visualpde.com/sim/?preset=waveEquation1DValidity} of the wave equation matched against the analytical solution in black. \textbf{(c)--(d)} \rev{An \hereVPDEfig[example of stripe to spot instabilities]{https://visualpde.com/sim/?preset=GiererMeinhardtStripeICs} in the Gierer--Meinhardt system}. \textbf{(e)--(f)} \hereVPDEfig[Image denoising]{https://visualpde.com/sim/?preset=PeronaMalik} via the Perona--Malik equation.}
    \label{fig:teaching_examples}
\end{figure}
\addtocounter{footnote}{-3}
\stepcounter{footnote}\footnotetext{\url{https://visualpde.com/sim/?preset=waveEquation1DValidity}}
\stepcounter{footnote}\footnotetext{\url{https://visualpde.com/sim/?preset=GiererMeinhardtStripeICs}}
\stepcounter{footnote}\footnotetext{\url{https://visualpde.com/sim/?preset=PeronaMalik}}

There are several different ways that users can share and extend the VisualPDE website and simulator, described in detail \rev{on the \hereVPDE[VisualPDE FAQs page]{https://visualpde.com/user-guide/FAQ.html\#sharing}}. These include sharing a screenshot, sharing a link to a user-crafted model, embedding a model inside another website, or forking the entire VisualPDE website or simulator via the GitHub code \citep{github}. We expect that most users will design a simulation, either by modifying an existing example or writing their own, and then share this via a direct link. This can be done from within the `share' menu, accessed by clicking the `share' icon on the right-hand side of the screen. This will generate a URL that corresponds exactly to the current simulation given the specified initial conditions (it will not include anything drawn with the brush, timesteps taken, or any changes to the image files $I_S$ or $I_T$). We expect that this functionality meets the requirements of the vast majority of users, including those creating bespoke models for teaching or research.

Individual simulations can be directly embedded within a webpage by clicking `Embed' within the share menu, which generates a fragment of HTML that can be pasted into a user's own site. The result of this can be seen in action in the `Visual Stories' collection on the website. One can even customise the level of UI elements which are displayed, depending on how they envision a user interacting with their model. We can see this being invaluable in designing webpages to describe models in more detail, such as in research pages or interactive lecture notes.

Lastly, the entire code base is shared through a standard CC BY open source licence, meaning that users are allowed to directly fork the GitHub repository and design their own versions of the website, or even the simulator. The website has been designed with this idea of extensibility in mind, as the vast majority of it is written using high-level markdown, so that creating collections of examples like those currently on the site can be done very easily. We imagine this detailed modification being used by educators to design bespoke courses or lecture notes with VisualPDE at their core.

\section{VisualPDE in teaching, research \& knowledge exchange}\label{sec:examples}

Here, we briefly outline a few current projects and activities involving VisualPDE in different settings, and discuss the possibilities of using it much more widely.

\subsection{Teaching through interactive simulations}
The systems presented in the basic PDEs section of the website include many of the classical linear PDEs studied in undergraduate courses, such as the heat and wave equations, in addition to relatively simple extensions including the convection--diffusion and Euler buckling models. These examples include demonstrations of the d'Alembert solution of the wave equation (see \cref{fig:teaching_examples}a--b), and the Fourier series solution of the heat equation. The overall goal is to provide some intuition for mathematical formulas obtained analytically (such as the role of the wavenumber in the decay of a cosine initial condition in the heat equation), as well as to go beyond what can be easily understood analytically, for instance exploring the impact of heterogeneous media on these simple models.

Many of the PDEs that appear in Murray's classical textbooks on mathematical biology are included on the website \citep{murray2003mathematical}. These books form the core of many mathematical biology courses, and cover a range of topics that can benefit from interactive visualisations. Two of the authors of this manuscript have begun making extensive use of VisualPDE in teaching a third year undergraduate mathematical biology course at Durham University,  providing links and interactive demonstrations of VisualPDE simulations in lectures and via the course's content management system. Initial informal feedback from students suggests that VisualPDE has significantly enhanced students' understanding of a range of topics in the course. These include travelling waves in the \hereVPDE[Fisher--Kolmogorov equation]{https://visualpde.com/mathematical-biology/travelling-wave.html}, impacts of Allee effects on \hereVPDE[spatial population invasion]{https://visualpde.com/mathematical-biology/bistable-travelling-waves.html}, and pattern formation and stripe vs spot selection in the \hereVPDE[Gierer--Meinhardt system]{https://visualpde.com/mathematical-biology/gierer-meinhardt.html}. 
 
Importantly, all of these topics illustrate theory that can be developed and understood with pen and paper, while also highlighting phenomena that are analytically intractable at this level but easy to explore numerically. For example, we encourage students to explore how the initial mass and shape of an invasive population matters for persistence when subject to a strong Allee effect, and compare this with a spatially homogeneous model where persistence is much easier to understand due to the simplicity of an equilibrium acting as a separatrix. \cref{fig:teaching_examples}c--d shows an example simulation in Gierer--Meinhardt demonstrating how stripe-like solutions are unstable to small perturbations, breaking up into spots. Such a result is difficult to understand analytically \citep{kolokolnikov2006stability}, but one can develop intuition through simulations.

Along with using VisualPDE to enhance engagement and build intuition in a lecture setting, we have been including interactive exercises on homework sheets that encourage students to numerically verify their analytical calculations of, for example, thresholds for Turing instability or wavespeeds for travelling wave models. In future, we hope to further capitalise on the flexibility of VisualPDE and provide students with opportunities to exhibit creative expression in their mathematical explorations, such as by designing their own biologically meaningful systems \citep{woolley2021bespoke}. In the broader context of undergraduate education, we hope that VisualPDE can be integrated into a wide range of courses and incorporated into a new generation of interactive course assessments.

\begin{figure}
    \centering
    (a) \includegraphics[width=0.27\textwidth,valign=c]{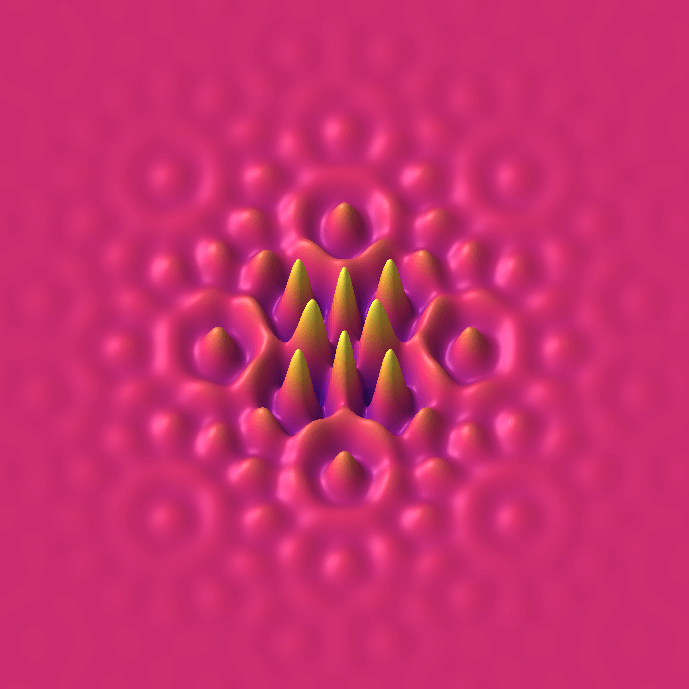}\, $\to$ \,
    (b) \includegraphics[width=0.27\textwidth,valign=c]{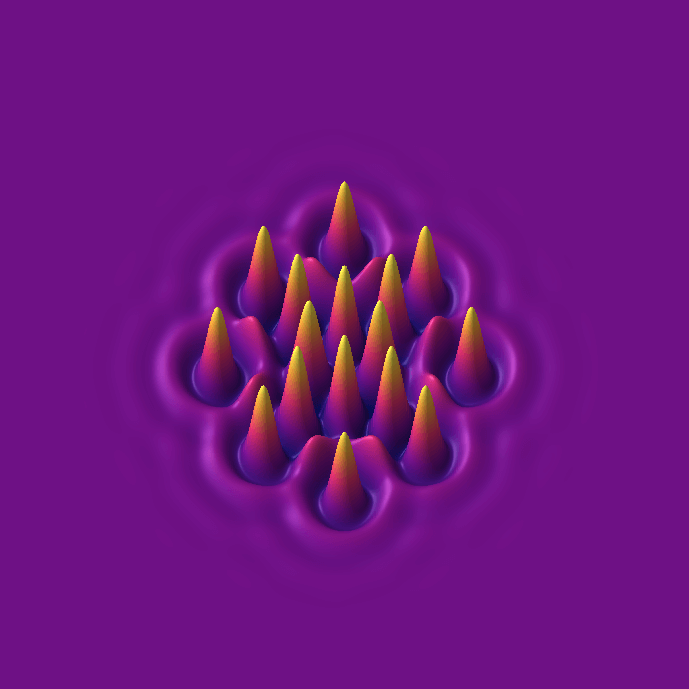}\, $\to$ \,
    (c) \includegraphics[width=0.27\textwidth,valign=c]{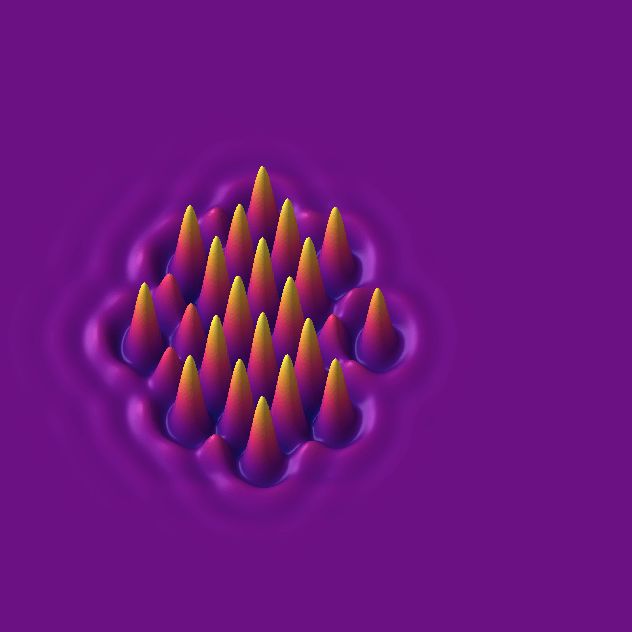}\\
    \vspace{5mm}
    (d) \includegraphics[width=0.27\textwidth,valign=c]{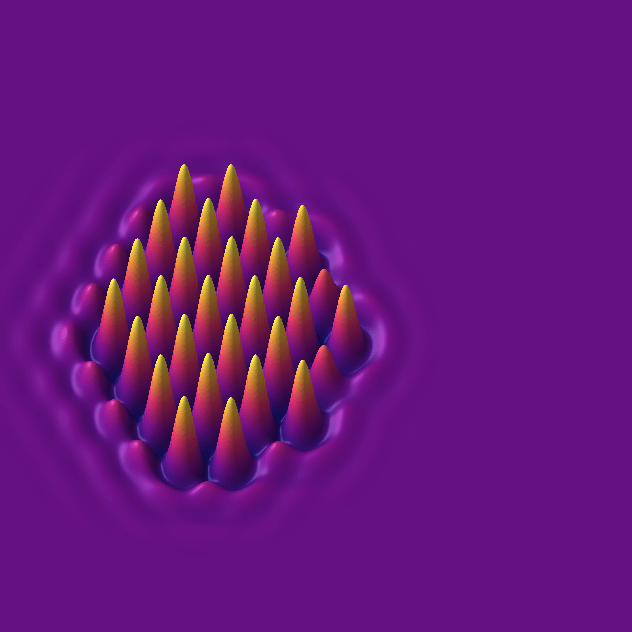}\, $\to$ \,
    (e) \includegraphics[width=0.27\textwidth,valign=c]{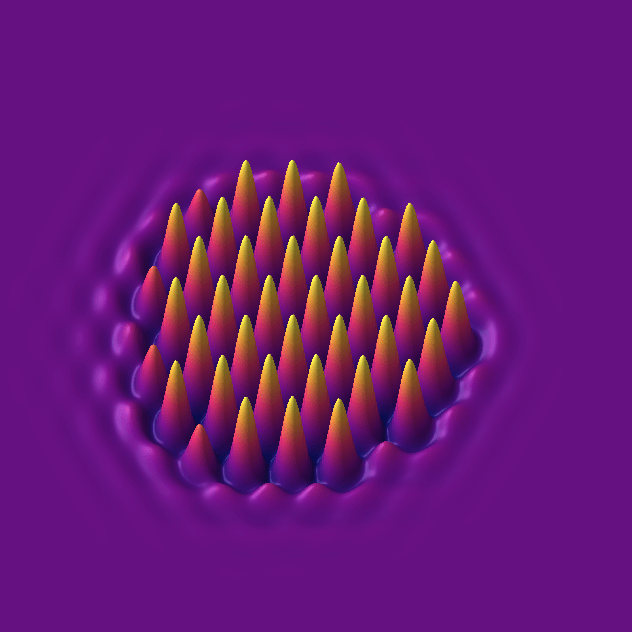}    \, $\to$ \,
    (f) \includegraphics[width=0.27\textwidth,valign=c]{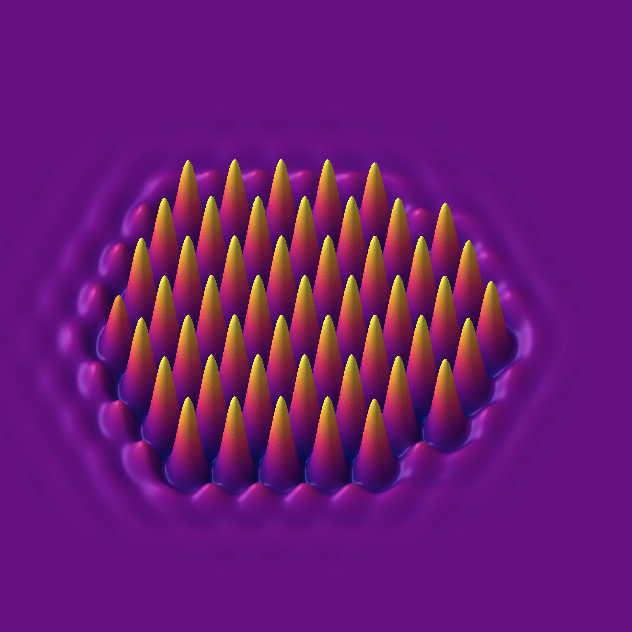}
    \caption{Example simulations of the Swift--Hohenberg equation giving rise to \hereVPDEfig[translating localised solutions]{https://visualpde.com/sim/?preset=swiftHohenbergRight}, which can be explored with VisualPDE. \textbf{(a)--(b)} Localised patterns in the form from an initial condition in (a) to the stationary structure with $D_4$ symmetry in (b). \textbf{(c)--(f)} Under sufficiently large advection to the right, the localised structure from (b) undergoes an instability forming a hexagonal pattern on its left side, breaking the $D_4$ symmetry. Captured July 2023 on a $1280\times768$px screen.}
    \label{fig:localised_examples}
\end{figure}
\footnotetext{\url{https://visualpde.com/sim/?preset=swiftHohenbergRight}}

Beyond introductory PDEs and mathematical biology, colleagues have reported using VisualPDE to give students access to visualisations in other topics. For example, the Korteweg--De Vries equation simulation mentioned in \cref{sec:numerical-methods} exhibits an important phenomenon, in which two solitons pass through one another without interacting \rev{(noting that this is only captured approximately, as described in \cref{sec:numerical-methods})}. The \hereVPDE[Perona--Malik equation]{https://visualpde.com/nonlinear-physics/perona-malik.html}, used for image denoising, particularly benefits from interactive exploration, as the ability of the model to sharpen an image can depend somewhat sensitively on the parameters (see \cref{fig:teaching_examples}e--f for a successful denoising simulation). These are just two examples of what we consider to be important lessons that can often be lost in pen-and-paper analysis or the details of numerical implementations, but which can be immediately appreciated by simply trying it out using different parameters and images in VisualPDE. As noted in \cref{sec:history}, we view such playing at the concept level as valuable scaffolding for in-depth courses in topics such as image analysis and numerical methods.

In the UK context, it is common for lecturers to use bespoke lecture notes for a given course. In the past few years it has become standard to make these notes available to students, and it is becoming increasingly desirable to make them accessible in HTML, which has several advantages compared to PDF, especially for students with disabilities \citep{mejia2021survey}. One advantage to web-based tools such as VisualPDE, GeoGebra and Desmos is that they can be natively embedded in such webpages, providing seamless connection with the material being described. There are currently several efforts to develop accessible and interactive lecture notes, and we are hopeful that VisualPDE can bring substantial value to these web-based formats. Of course, there are important challenges regarding accessibility of such visual tools themselves, and this is an ongoing area of current work.

\subsection{VisualPDE in research development and communication}
\begin{figure}
    \centering 
    (a) \includegraphics[width=0.45\textwidth,valign=c]{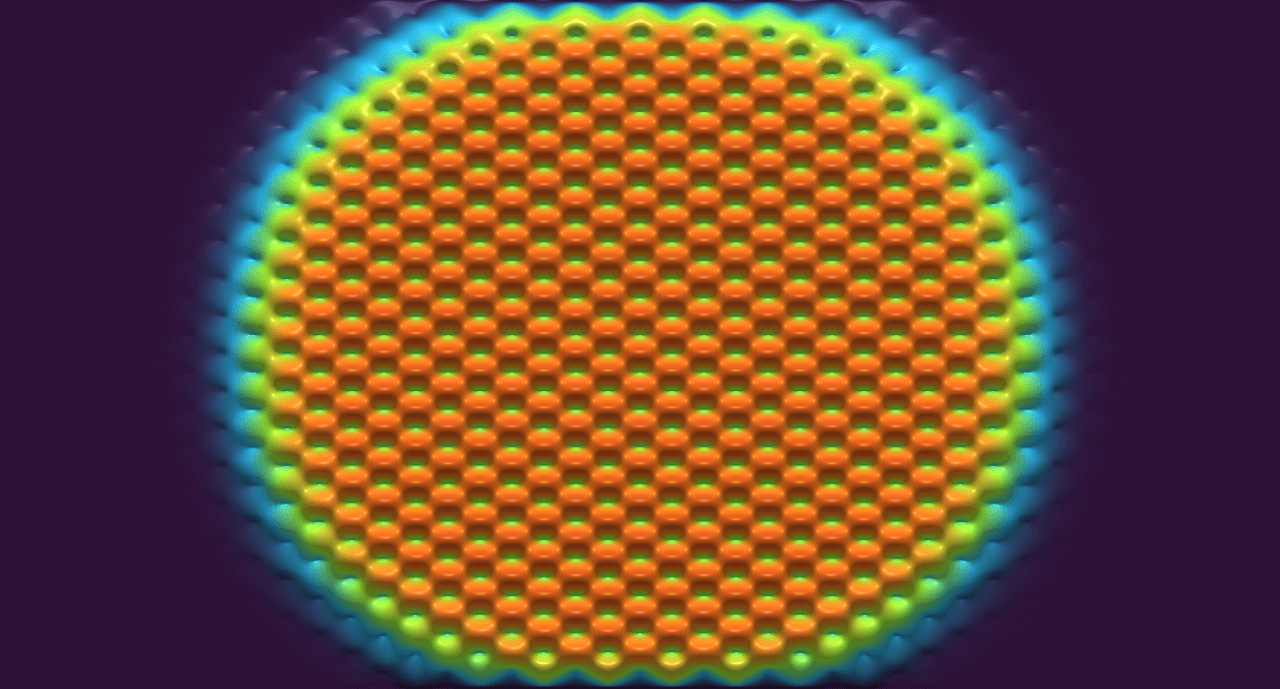}
    (b) \includegraphics[width=0.45\textwidth,valign=c]{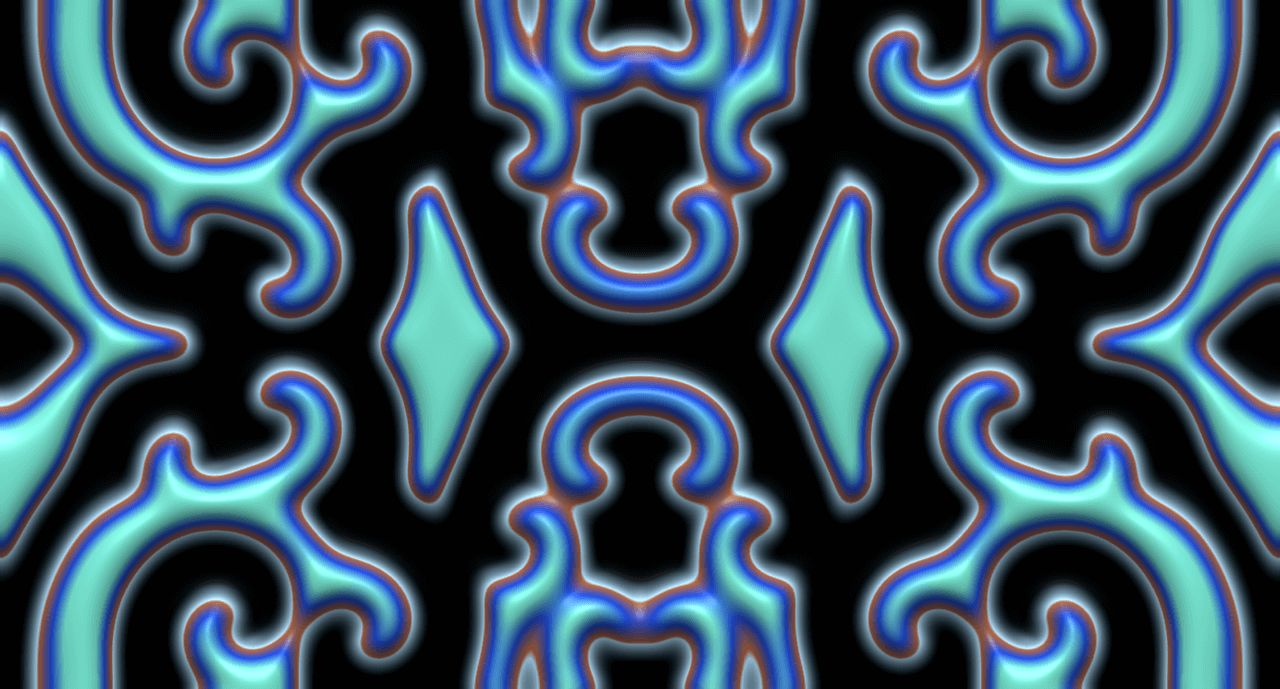}\\
    \vspace{5mm}
    (c) \includegraphics[width=0.45\textwidth,valign=c]{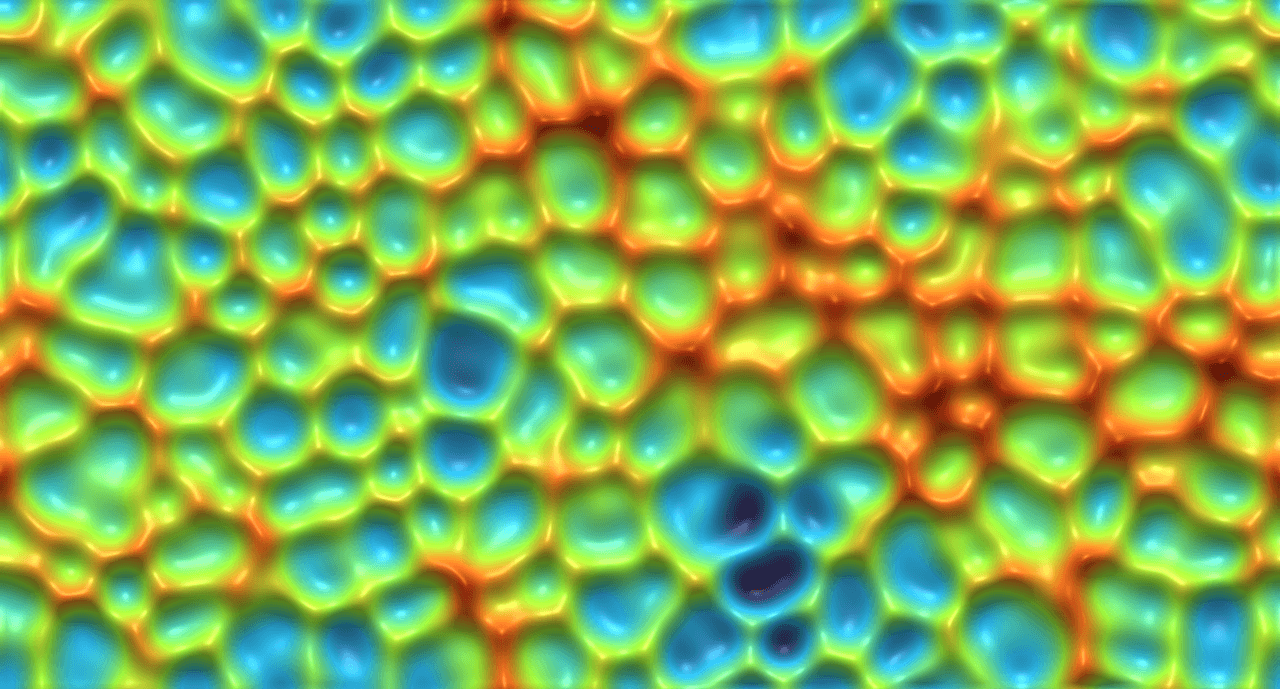}
    (d) \includegraphics[width=0.45\textwidth,valign=c]{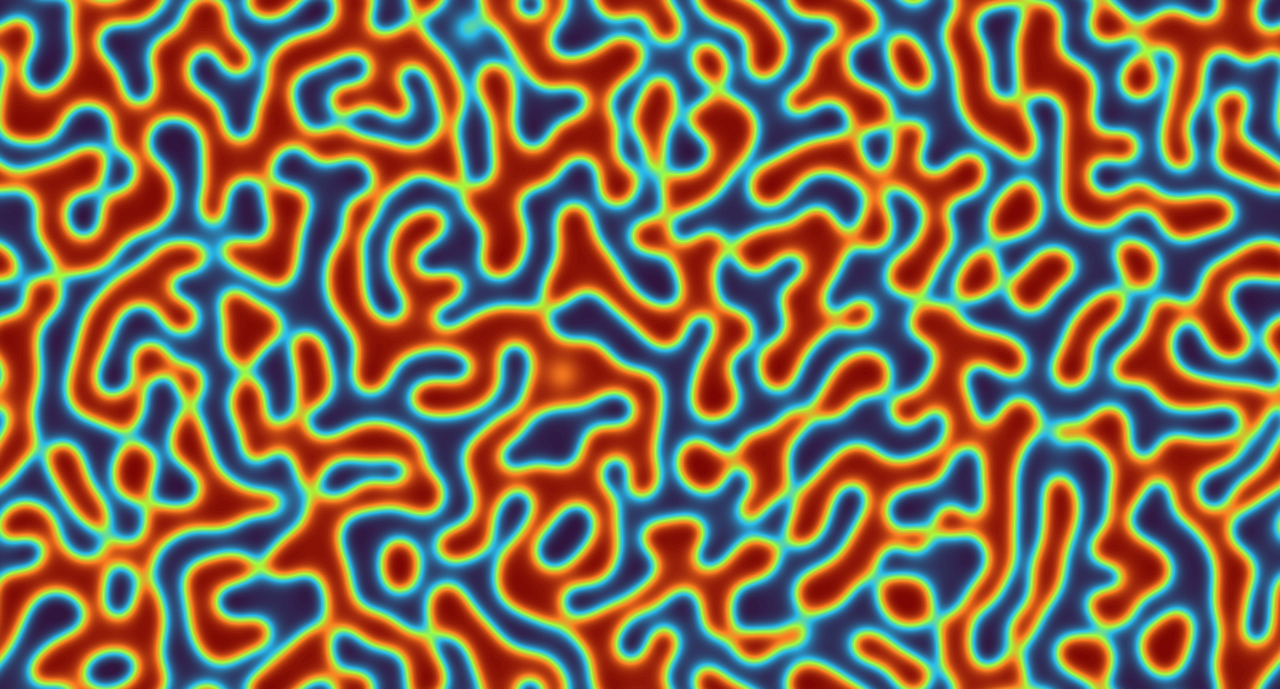}\\
    \vspace{5mm}
    (e) \includegraphics[width=0.45\textwidth,valign=c]{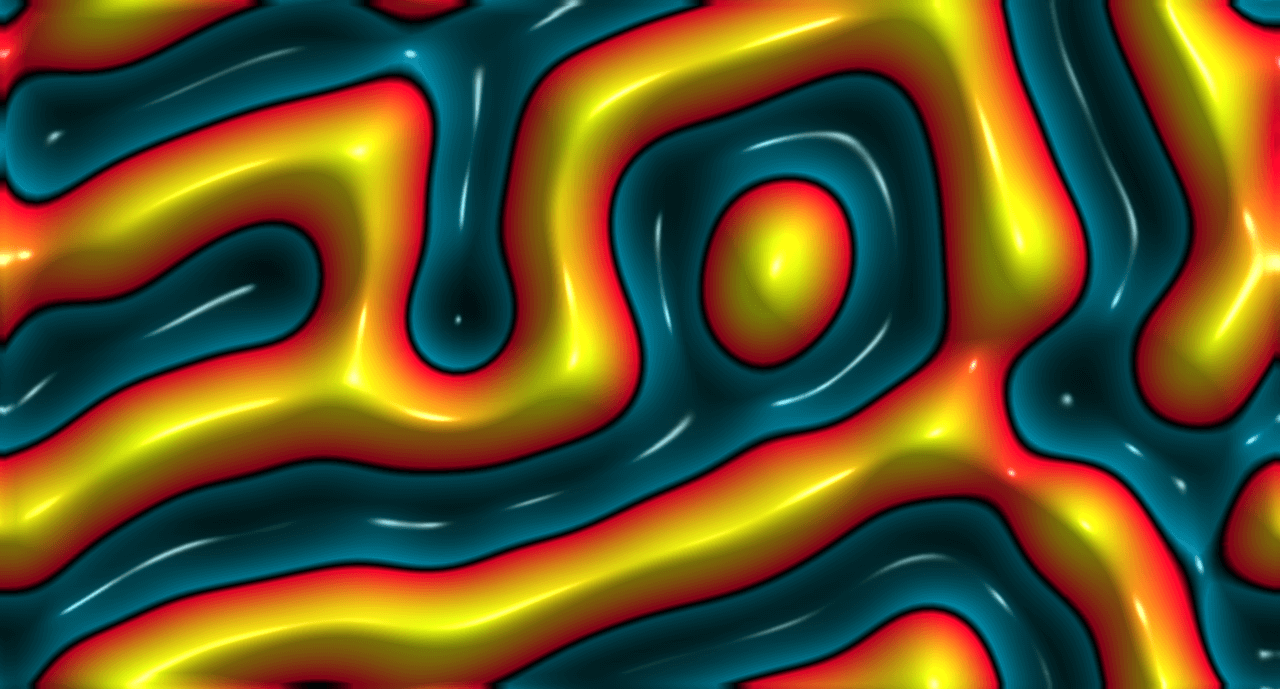}
    (f) \includegraphics[width=0.45\textwidth,valign=c]{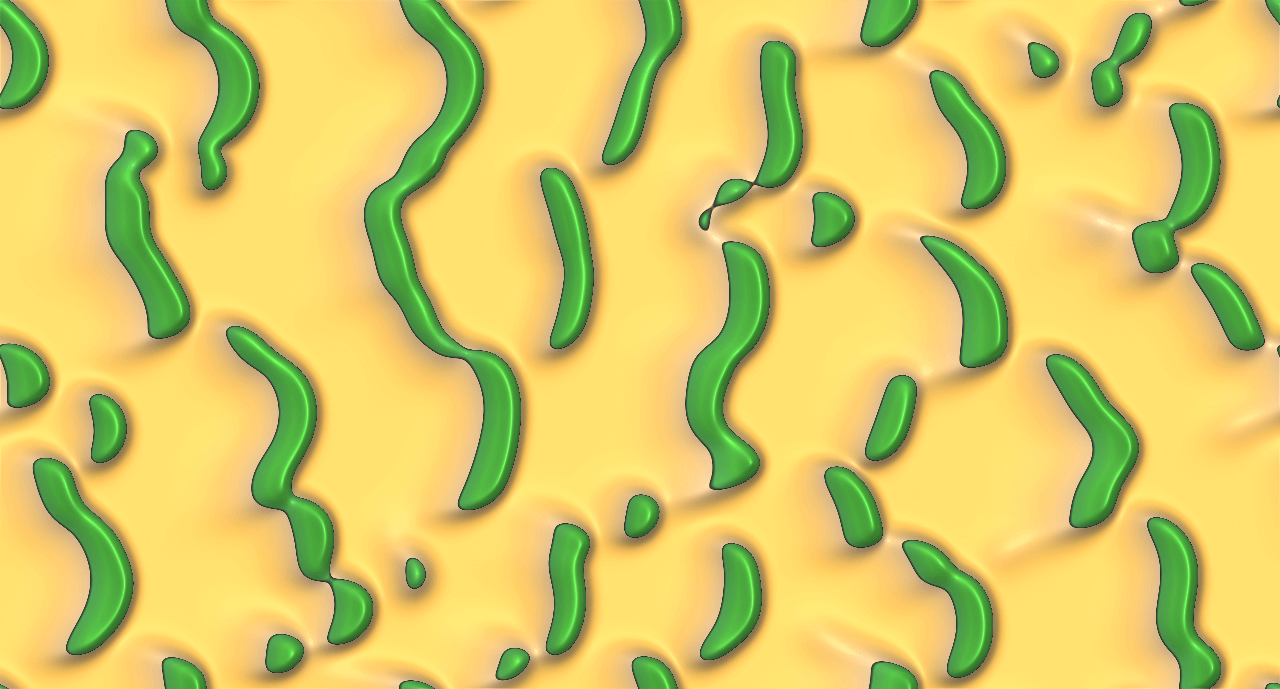}
    \caption{Example simulations of spatiotemporal solutions. \textbf{(a)} A \hereVPDEfig[heterogeneous Fisher--Kolmogorov equation]{https://visualpde.com/sim/?preset=InhomogeneousFisherKPP}. \textbf{(b)} A \hereVPDEfig[three-species Lotka--Volterra system]{https://visualpde.com/sim/?preset=cyclicCompetition}. \textbf{(c)} The \hereVPDEfig[Kuramoto–Sivashinsky equation]{https://visualpde.com/sim/?preset=KuramotoSivashinsky}. \textbf{(d)} An example of \hereVPDEfig[coarsening in the Cahn--Hilliard equation]{https://visualpde.com/sim/?preset=CahnHilliard}. We recommend playing with the timescale $r$ in this simulation.
    \textbf{(e)} \hereVPDEfig[Turing wave bifurcations]{https://visualpde.com/sim/?preset=TuringWaveFHN2D} in a hyperbolic reaction--diffusion system. 
    \textbf{(f)} Irregular vegetation patterns in the \hereVPDEfig[Klausmeier model]{https://visualpde.com/sim/?preset=KlausmeierModel}. Captured July 2023 on a $1280\times768$px screen.}
    \label{fig:spatiotemporal_examples}
\end{figure}
\addtocounter{footnote}{-6}
\stepcounter{footnote}\footnotetext{\url{https://visualpde.com/sim/?preset=InhomogeneousFisherKPP}}
\stepcounter{footnote}\footnotetext{\url{https://visualpde.com/sim/?preset=cyclicCompetition}}
\stepcounter{footnote}\footnotetext{\url{https://visualpde.com/sim/?preset=KuramotoSivashinsky}}
\stepcounter{footnote}\footnotetext{\url{https://visualpde.com/sim/?preset=CahnHilliard}}
\stepcounter{footnote}\footnotetext{\url{https://visualpde.com/sim/?preset=TuringWaveFHN2D}}
\stepcounter{footnote}\footnotetext{\url{https://visualpde.com/sim/?preset=KlausmeierModel}}

This project was initially conceptualised in terms of teaching. However, we think there is ample potential to use it for rapid prototyping and communication of a variety of spatial models, and we are presently using it in our own research. Below we showcase some examples of this where we discuss the ease with which spatial models can be developed and shared, and highlight the potential for VisualPDE to drastically alter how we communicate and interact with mathematics during the research cycle.

\subsubsection{Spatially heterogeneous reaction--diffusion dynamics} 
\citet{page2003pattern,page2005complex} explored models of spatially heterogeneous reaction--diffusion systems finding, among other things, that such heterogeneity can lead to spatiotemporal movement of spike solutions in 1D. This dynamic behaviour was later explored by \citet{krause2018heterogeneity} and \citet{kolokolnikov2018pattern} using different methods.  Related to these studies, \citet{dillon1999mathematical} explored mixed boundary conditions using continuation methods, and \cite{krause2020isolating} later justified these conditions in terms of asymptotic models of heterogeneous tissue, finding that certain combinations of mixed boundary conditions can isolate patterning away from the boundary of the domain. In all of these cases, the phenomenon was largely understood first by numerical exploration before any analytical results were available. VisualPDE allows for rapid implementations of every model studied in these papers, such as this \hereVPDE[example of dynamic bifurcations with heterogeneity]{https://visualpde.com/mathematical-biology/heterogeneous-dynamics.html} , including explorations of the dynamic spatiotemporal spike oscillations observed first by \cite{page2005complex}. While our implementation of these models in VisualPDE is some years after these papers were written, one could imagine developing these ideas substantially more quickly through the rapid prototyping and analysis made possible by VisualPDE.

\subsubsection{Localised structures} 
Localised patterns are steady states of PDEs with spatial structure only in a subset of the domain, with most of the domain at a homogeneous and stable steady state. They represent an interesting class of multistable solutions in PDEs, and substantial recent research has been undertaken to understand them \citep{burke2007homoclinic, knobloch2011isolas, champneys2021bistability}. Recently, \cite{hill2023approximate} explored a class of symmetric localised solutions in two spatial dimensions in the Swift--Hohenberg equation. The lead author of the cited study was able to quickly use these results to develop a VisualPDE model capable of generating several classes of two-dimensional solutions exhibiting different symmetries, one of which is shown in \cref{fig:localised_examples}b and can be \hereVPDE[explored on VisualPDE]{https://visualpde.com/nonlinear-physics/swift-hohenberg.html}. Once these states are found numerically within VisualPDE, it is easy to subject them to various perturbations, both in terms of perturbing the solution state but also in terms of changing the model. For instance, an \hereVPDE[example exploring advection]{https://visualpde.com/nonlinear-physics/advecting-patterns.html} considers the effect of rotational and linear advection on these localised states. For small values of advection the states persist with minor changes but, as the advective velocity increases, they begin to change shape and lose symmetries. An example is shown in \cref{fig:localised_examples}c--f, where a large rightward advection has changed the shape of the localised structure, with it developing a larger hexagonally-spaced `tail'. Importantly, the solution no longer retains the $D_4$ symmetry used to rigorously study its existence and stability. This \rev{loss of symmetry} immediately leads to questions of how to extend such analyses to consider more complicated settings where the originally assumed symmetries are not preserved in the model and, more broadly, to understand how robust such solutions are to changes in the model. Notably, extending this model within VisualPDE to include advection took only a matter of seconds, significantly less than the time needed to develop a numerical scheme to add advection in a programming language. 

\subsubsection{Spatiotemporal dynamics}
Another area in which rapid and interactive PDE simulations are extremely valuable is the domain of spatiotemporal solutions, such as solitons, travelling and spiral waves, and chaos. For such systems, 2D snapshots are often insufficient to really gain intuition about the dynamics compared to moving videos. Many such systems are also somewhat sensitive to initial conditions and other details of the model, so being able to tweak these interactively allows for a deeper understanding of the roles of such details in the resulting dynamics. \Cref{fig:spatiotemporal_examples} presents some exemplar systems, including invasion in a spatially heterogeneous Fisher--Kolmogorov equation, spiral waves in a three-species `rock--paper--scissors' Lotka--Volterra system \citep{reichenbach2007mobility}, spatiotemporal chaos in the Kuramoto–Sivashinsky equation \citep{kalogirou2015depth}, coarsening in the Cahn--Hilliard equation \citep{miranville2019cahn}, Turing wave bifurcations in hyperbolic reaction--diffusion systems \citep{ritchie_hyperbolic_2020}, and irregular vegetation patterns in the Klausmeier model \citep{klausmeier1999regular}. These examples have many connections to contemporary research questions where we feel the lightning-fast interactivity of VisualPDE can be helpful in stimulating discussion and understanding broad qualitative features of different models.

\subsubsection{Research communication} 
Besides the examples above, the authors have also found the use of VisualPDE in interacting with collaborators, colleagues and industrial partners invaluable. Being able to rapidly test an idea and share a link directly to a simulation (with all parameters and functional forms easily accessible) has been greatly helpful in speeding up our own research efforts. A use case that we envisage becoming more and more common is for demonstrating prototypes to industrial and interdisciplinary partners, with VisualPDE enabling solution and visualisation within minutes of a model first being conceived. Even in reviewing papers, we have found the ability to quickly corroborate qualitative results useful. We suspect that as tools like VisualPDE become more widespread, it will become much more routine to send direct links to open-source implementations of models at high levels. We also hope that, as in this article, it will become commonplace to include links to interactive simulations in research articles, allowing the research community to engage with published science in an entirely new way.

\subsection{Knowledge exchange through `Visual Stories'}

\begin{figure}
    \centering 
    (a) \includegraphics[width=0.4\textwidth,valign=c]{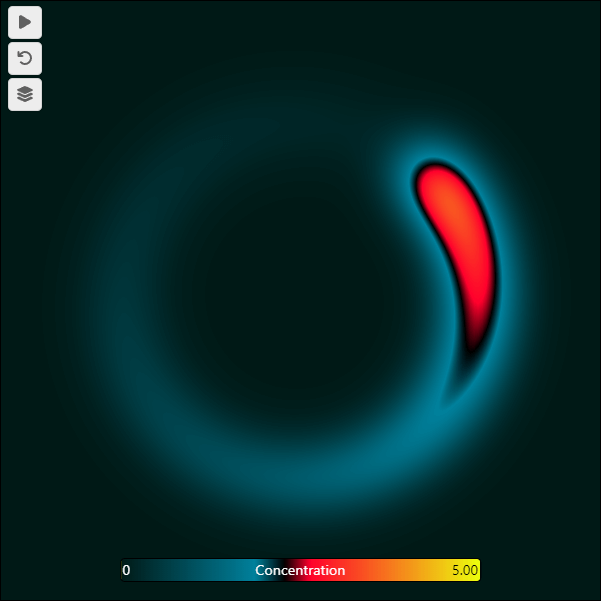}
    (b) \includegraphics[width=0.4\textwidth,valign=c]{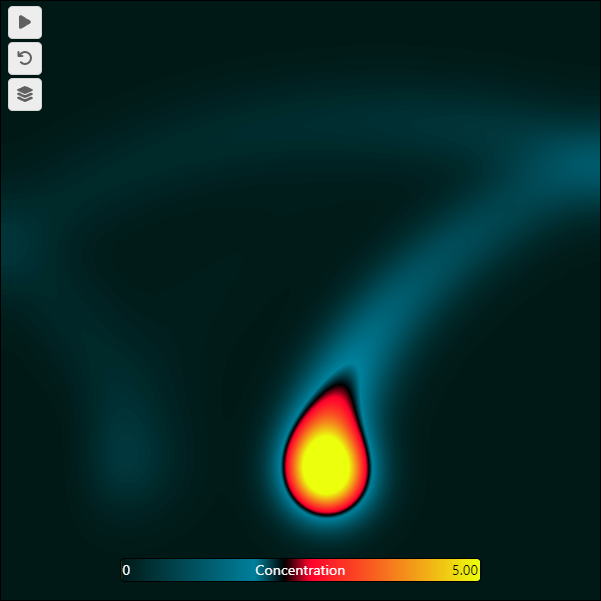} 
    \caption{Example simulations of the concentration of virus particles from a rotating source, from an adaptation of the model of \citet{lau2022predicting}. \textbf{(a)} A circling source only. \textbf{(b)} A circling source subject to advection to the right due to an air conditioner. See the \href{https://visualpde.com/visual-stories/airborne-infections.html}{Visual Story}\cref{footnote: covid story} for more details.}
    \label{fig:Covid_story}
\end{figure}

Beyond teaching or research, we think VisualPDE has marked potential for broad knowledge exchange and outreach activities. We have begun developing a `maths-free' collection of examples that focuses on models as ways to explore phenomena, doing so purely through interactive simulation with VisualPDE and a guiding narrative aimed at the general public. These can be found in the \hereVPDE[`Visual Stories' collection]{https://visualpde.com/visual-stories.html} on the website.

One example of the use of VisualPDE for knowledge exchange is in exploring airborne virus transmission within a room, incorporating the effects of circulating airflow. \citet{lau2022predicting} developed a model of this situation, which eventually led to significant interaction with policymakers during the Covid-19 pandemic and the development of a web-based airborne virus risk calculator\footnote{\url{https://people.maths.ox.ac.uk/griffit4/Airborne_Transmission/index.html}}. With support of the authors of the cited study, we extended this interface using VisualPDE to allow for more detailed spatial probabilities of infection in a Visual Story on virus transmission. This innovative medium complements the existing calculator by providing an interactive way for diverse audiences to engage with the various features of the model and develop their own intuition for complex mathematics. See \cref{fig:Covid_story} for some snapshots of this Story, or interact with \hereVPDELabel[the Story]{https://visualpde.com/visual-stories/airborne-infections.html}{footnote: covid story} yourself .

Notably, creating this accessible exposition of complex, cutting-edge mathematics was straightforward using VisualPDE. In particular, producing visually striking simulations that are woven into the narrative is simple, with interactive elements being embedded directly into the article. We hope that this inspires other researchers to explore this medium as a way to communicate their research and knowledge to a broad audience, with engagement and interaction at the heart of the discourse.

\section{Summary}\label{sec:summary}
In this paper, we have presented a web-based interactive PDE simulator capable of solving a large range of time-dependent PDEs in real time. We have discussed the historical and pedagogical context behind this tool, its technical and user-facing design philosophy, and examples of its use in teaching, research, and knowledge exchange, which we hope illustrate its broad, multi-audience potential. Looking ahead, we hope that VisualPDE proves to be useful across many mathematical communities, and that it facilitates and inspires interdisciplinary connections through interactive, accessible, shareable computing. We also hope that, alongside tailored numerical codes, future research articles across disciplines might include links to interactive, representative simulations like those presented in this article, engaging audiences in what we believe is an entirely new way. We are excited to continue developing VisualPDE and exploring its potential for changing how we interact with and communicate science and mathematics. We are hopeful that VisualPDE inspires further efforts in making mathematics more interactive, which we see as an exciting and important frontier in how we communicate and conceptualise science.

\begin{acknowledgements}
The ideas for this project originated in a Durham Centre for Academic Development collaborative innovation grant titled \emph{Accessible interactive visualisations in mathematical biology}, which supported AKC in the initial version of the interactive PDE solver, based on the Gray--Scott reaction--diffusion simulator by \citet{pmneila2012grayscott}. BJW is supported by the Royal Commission for the Exhibition of 1851.
\\
\item[\hskip\labelsep
\bfseries{Data Availability}]
There is no data present in this paper. All code associated with the project can be found on GitHub \citep{github}.
\end{acknowledgements}

\bibliographystyle{abbrvnat}
\bibliography{refs}

\end{document}